\documentclass[aps,prb]{revtex4}
\usepackage{graphicx}
\usepackage{epsfig}

\begin{document}

\title{Applications of exact solution for strongly interacting one dimensional bose-fermi mixture: low-temperature correlation functions, density profiles and collective modes.}

\author{Adilet Imambekov\footnote{email address imambek@cmt.harvard.edu} and Eugene Demler}
\affiliation{Department of Physics, Harvard University, Cambridge
MA 02138}

\date{\today}

\begin{abstract}
We consider one dimensional interacting bose-fermi mixture with
equal masses of bosons and fermions, and with equal and repulsive
interactions between bose-fermi and bose-bose particles.
Such a system can be realized in  current experiments with ultracold bose-fermi mixtures.
We apply the Bethe-ansatz technique to find the exact ground state energy at
zero temperature for any value of interaction strength and density
ratio between bosons and fermions. We use it to prove the absence of the demixing,  contrary to  prediction of a mean field approximation. Combining exact solution with local density approximation (LDA) in a harmonic trap, we calculate the density profiles and frequencies of collective modes in various limits. In the strongly interacting regime, we predict the appearance of low-lying
collective oscillations which correspond to the counterflow of the two species. In the strongly interacting regime we use exact wavefunction to calculate the single particle correlation functions for bosons and fermions at low temperatures under periodic boundary conditions. Fourier transform of the correlation function is a momentum distribution, which can be measured in time-of-flight experiments or using Bragg scattering. We derive an analytical formula, which allows to calculate correlation functions at all distances numerically for a polynomial time in the system size. We investigate numerically two strong singularities of the momentum distribution for fermions at $k_f$ and $k_f+2k_b.$ We show, that in strongly interacting regime correlation functions change dramatically as temperature changes from $0$ to a small temperature $\sim E_f/\gamma \ll E_f,$ where $E_f=(\pi \hbar n)^2/(2m), \; n$ is the total density and $\gamma =mg/(\hbar^2n) \gg 1$ is the Lieb-Liniger parameter. A strong change of the
momentum distribution in a small range of temperatures can be used to perform a thermometry
at very small temperatures.

\end{abstract}
\maketitle

\section{Introduction}
\label{introduction}
Recent developments in cooling and trapping of cold atoms open exciting opportunities
for experimental studies of interacting systems under well controlled conditions. Current experiments \cite{bfexp,fermipressure} can deal not only with single component gases, but with various atomic mixtures. Using Feshbach\cite{KRbFeshbach,LiNaFeshbach} resonances and and/or optical
lattices\cite{Jaksch98,Bloch}  one can tune different parameters, and drive the systems towards strongly correlated regime.  The effect of correlations is most prominent for low
dimensional systems, and recent experimental
realization\cite{Weiss,Paredes} of a strongly interacting
Tonks-Girardeau (TG) gas of bosons opens new perspectives in
experimental studies of strongly interacting systems in
1D\cite{Moritz1dmolecules}. In this article we investigate bose-fermi mixtures in 1D,
using exact techniques of the Bethe ansatz. Some of the results presented here
have been reported earlier \cite{bfshort}.

Most of the theoretical research on bose-fermi mixtures\cite{bftheory} so far has been concentrated on higher dimensional systems, and only recently 1D systems started attracting attention. Several properties of such systems have been investigated so far, including
phase separation\cite{Das,CazalillaHo,jap_numerics}, fermion pairing\cite{Mathey},
possibility of charge density wave (CDW) formation\cite{CDW} and long distance
behavior of correlation functions\cite{Frahm}.

  A 1D interacting bose-fermi mixture is described by the Hamiltonian
\begin{equation}
H=\int_0^L dx (\frac{\hbar^2}{2m_b}\partial_x \Psi_b^\dagger\partial_x
\Psi_b+\frac{\hbar^2}{2m_f}\partial_x \Psi_f^\dagger\partial_x \Psi_f) +
\int_0^L dx (\frac12g_{bb}\Psi_b^\dagger
\Psi_b^\dagger\Psi_b\Psi_b + g_{bf} \Psi_b^\dagger
\Psi_f^\dagger\Psi_f\Psi_b ). \label{initialhamiltoniansc}
\end{equation}
Here, $\Psi_b, \Psi_f$ are boson and fermion operators, $m_b, m_f$
are the masses, and $g_{bb}, g_{bf}$ are bose-bose  and bose-fermi
interaction strengths. The model (\ref{initialhamiltoniansc}) is exactly solvable, when\cite{Lai}
\begin{eqnarray}
m_f=m_b=m, \,  g_{bb}=g_{bf}=g >0.\label{intercond}
\end{eqnarray}
 It corresponds to the situation when masses are the same,
and bose-bose and bose-fermi interaction strengths are the same
and positive.  Although conditions (\ref{intercond}) are somewhat restrictive,
 the exactly solvable case is relevant to current experiments (the experimental situation will be analyzed in detail in section \ref{experiments} ) and
 can be used to check the validity of different approximate approaches.
 Model (\ref{initialhamiltoniansc}) under conditions (\ref{intercond}) has been considered in
 the literature before\cite{Lai}, but its properties have not been investigated in detail. After the appearance of our initial report \cite{bfshort}, two additional articles \cite{Frahm,Batchelor} used Bethe ansatz to investigate the same model.
 We use the exact solution to calculate the ground state energy and investigate phase separation and collective modes at zero temperature.
 For strongly interacting regime, we calculate single particle correlation functions,
 and consider the effects of small temperature on correlation functions
 and density profiles.

 The article is organized as follows. In section \ref{BAreview}
 we review the Bethe ansatz solution  for bose-fermi mixture and compare it to the solution for fermi mixture. In section \ref{numerics} we obtain the energy numerically in the thermodynamic limit. We use it to prove the absence of the demixing under conditions (\ref{intercond}), contrary to  prediction of a  mean field \cite{Das} approximation. In section \ref{LDAsection} we combine exact solution with local density approximation (LDA) in a harmonic trap, and calculate the density profiles and frequencies of collective modes in various limits. In the strongly
interacting regime, we predict the appearance of low-lying
collective oscillations which correspond to the counterflow of the
two species. In section \ref{zerotcorrfunc} we use exact wavefunction in the strongly interacting regime to calculate the single particle correlation functions for bosons and fermions at zero temperature under periodic boundary conditions.
We derive an analytical formula, which allows to calculate correlation functions at all distances numerically for a polynomial time in system size.
In section \ref{nztcorrelations} we extend the results of section \ref{zerotcorrfunc} for low temperatures. We also calculate the evolution of the  zero temperature density profile
at small nonzero temperatures. We show, that in strongly interacting regime correlation functions change dramatically as temperature is raised from $0$ to a small value.
Finally in section \ref{experiments} we analyze the experimental situation and make concluding remarks.

\section{Bethe ansatz solution}
\label{BAreview}
In this section we will briefly review the solution \cite{Lai} of
the model (\ref{initialhamiltoniansc}) under periodic boundary conditions
and compare it to the
solution of Yang of the spin-$\frac12$ interacting fermions \cite{Yang67, Gaudin}, for
the sake of completeness. More details on Yang's solution can be
found in \cite{babooks,takahashi,Tsvelick,Andrei,Sutherlandbook}.

 In first quantization, hamiltonian (\ref{initialhamiltoniansc}) can be written as
\begin{equation}
H=-\sum_{i=1}^{N}\frac{\partial ^2}{\partial x_i ^2}+ 2c\sum_{i<j}
\delta(x_i-x_j),\; c >0. \label{initialhamiltonianfc}
\end{equation}
Here we have assumed $m=1/2$ and $\hbar=1,$ to keep contact with the
literature on the subject. Later in the discussion of the collective modes we
will introduce the mass of atoms, but it should be clear from the context whether
we have assumed $m=1/2$ or not. $c$ in (\ref{initialhamiltonianfc}) is connected
to parameters of (\ref{initialhamiltoniansc}) via
\begin{equation}
c=\frac{m g}{\hbar^2}.
\end{equation}

Wave  function is supposed to be symmetric with respect to
indices $i=\{1, ... ,M\}$ (bosons) and antisymmetric with respect
to $i=\{M+1,...,N\}$ (fermions). On the first stage, Yang's
solution doesn't impose any symmetry constraint on the
wavefunction. On the second stage, periodic boundary conditions
are resolved with the help of extra Bethe-ansatz. This idea has
been generalized by Sutherland \cite{Sutherland68} for the case of
N-fermion species. The results presented here can be simply
derived from Sutherland's work.

In Yang's solution, one assumes the generalized coordinate Bethe wavefunction of the following form:
for $0<x_{Q_1}<x_{Q_2}< ...< x_{Q_N}<L$
\begin{eqnarray}
\Psi=\sum_P [Q,P] e^{i\sum{k_{P_i} x_{Q_i}}},
E=\sum_i k^2_i.
\end{eqnarray}
where $k_1, ... ,k_N$ is a set of unequal numbers,  $P$ is an
arbitrary permutation from $S_N$ and $[Q,P]$ is $N!\times N!$
matrix. Let's denote the columns of this matrix as  $N!$
dimensional vector $\xi_P.$
 Delta function potential in (\ref{initialhamiltonianfc}) is equivalent to the following
boundary condition for the derivatives of the wavefunction:
\begin{equation}
(\frac{\partial}{\partial x_j}-\frac{\partial}{\partial x_k})\Psi_{x_j=x_k+0}- (\frac{\partial}{\partial x_j}-\frac{\partial}{\partial x_k})\Psi_{x_j=x_k-0}= 2 c \Psi_{x_j=x_k},
\label{bc}
\end{equation}
and the continuity condition reads
\begin{equation}
\Psi_{x_j=x_k+0}=\Psi_{x_j=x_k-0}.
\label{bc2}
\end{equation}

Suppose $Q$ and $Q'$ are two permutations, such that $Q_k=Q'_k,$
for $k \neq \{i, i+1\}, $ and $Q_i=Q'_{i+1},Q'_i=Q_{i+1}.$
Similarly, $P$ and $P'$ are two permutations, such that
$P_k=P'_k,$ for $k \neq \{i, i+1\}, $ and
$P_i=P'_{i+1},P'_i=P_{i+1}.$ To satisfy  (\ref{bc}) and
(\ref{bc2})  for $x_{Q_i}=x_{Q_{i+1}}$ independently of other $x,$
one has to impose two conditions for
 four coefficients $[Q,P],[Q',P],[Q,P'],[Q',P'].$
Using these two conditions, we can express $[Q,P'],[Q',P']$ via $[Q,P],[Q',P].$ These requirements can be simply written as
a condition between $\xi_P$ and $\xi_{P'}:$
\begin{equation}
\xi_{P'}=Y^{i,i+1}_{P_i,P_{i+1}} \xi_P.
\label{transfereq}
\end{equation}
$Y$ operators are defined as
\begin{equation}
Y^{l,m}_{i,j}=-\frac{\lambda_{ij}}{1+\lambda_{ij}}+\frac{1}{1+\lambda_{ij}}
\hat  P_{lm}, \label{Yangm}
\end{equation}
where $$ \lambda_{ij}= \frac{ic}{k_i -k_j}, $$ and $ \hat P_{lm}$
is an operator acting on a vector $\xi_P$ which interchanges the
elements with indices $Q_l$ and $Q_m.$ Using $Y$ operators one can
express any $\xi_P$ via $\xi_0,$ where $\xi_0$ is a column for
$P=$ identity. However, arbitrary permutation $P$ can be
represented as a combination of neighboring transpositions by
different means. Independence of the final result on a particular
choice of neighboring transpositions can be checked based on the
following Yang-Baxter Relations:
\begin{eqnarray}
Y^{a,b}_{i,j} Y^{a,b}_{j,i}=1, \\
Y^{a,b}_{j,k} Y^{b,c}_{i,k} Y^{a,b}_{i,j}=Y^{b,c}_{i,j} Y^{a,b}_{i,k} Y^{b,c}_{j,k}.
\label{YB}
\end{eqnarray}
Operators $Y^{i,i+1}_{P_i,P_{i+1}}$  exchange the momentum labels
$P_i$ and $P_{i+1},$ while $ \hat P_{i,i+1}$ interchange relative
position labels
 $Q_i$ and $Q_{i+1}$.  It is convenient to define combined  operator, which exchanges both labels:
 \begin{equation}
 X_{ij}=\hat P_{ij}Y^{ij}_{ij}=\frac{1- \lambda_{ij} \hat P_{ij}}{1+\lambda_{ij}}.
 \label{X}
 \end{equation}
 Using this definition, periodic boundary conditions can be written as $N$ matrix eigenvalue  equations:
 \begin{equation}
  X_{j+1,j}X_{j+2,j}...X_{N,j}X_{1,j}...X_{j-1,j}\xi_0=e^{i k_j L}\xi_0.
 \label{PBC}
 \end{equation}
 The procedure outlined above reduces equations  for $N!\times N!$ coefficients to  $N$ eigenvalue equations for $N!$ dimensional vector.
 Imposing some symmetry on $\xi_0$ simplifies the system further. If $\xi_0$ is antisymmetric with respect to particle permutations (fermions) ,
 then $\hat P_{ij}=-1$ and $e^{i k_j L}=1.$ The system of equations is the same as for noninteracting fermions, as expected. If $\xi_0$ is symmetric (bosons),
 $\hat P_{ij}=1$ and the system is equivalent to periodic boundary conditions of Lieb-Liniger model \cite{LL}.

 If one needs to consider two-species system, $\xi_0$ has the symmetry of the corresponding permutation group representation (Young tableau).
 Instead of solving eq. (\ref{PBC}), it is convenient to consider the similar problem in the conjugate representation. If $\xi_0$
 is antisymmetric with respect to  both permutations of the first $M$ indices and  the rest $N-M$ (two-species fermions),
 eigenstate in conjugate representation $\varphi$  is symmetric with respect to first $M$ indices and is  also symmetric with respect to permutations of the
  rest $N-M$ indices.
 Similarly, in conjugate representation for bose-fermi mixture with $M$ bosons and $N-M$ fermions $\varphi$
 should be chosen to be antisymmetric for permutations of $M$ boson indices and symmetric with respect to permutations of $N-M$ fermion indices.
   The periodic boundary conditions are (note the change of the sign in the definition of $X'_{ij}$ compared to $X_{ij}$):
 \begin{eqnarray}
  X'_{j+1,j}X'_{j+2,j}...X'_{N,j}X'_{1,j}...X'_{j-1,j} \varphi=e^{i k_j L}\varphi, \label{PBC21}\\
  X'_{ij}=\frac{1+ \lambda_{ij} \hat P_{ij}}{1+\lambda_{ij}}.
 \label{PBC22}
 \end{eqnarray}
  Since $N!$-dimensional vector $\varphi$ has symmetry constraints, it has  $C^M_N$ inequivalent components, characterized by the
  positions  $y_i$ of $M$ spin-down fermions (or $M$ bosons respectively).
  One can think of the components of the vector $\varphi$ as of
  the values of the spin wavefunction, defined on an auxiliary  one-dimensional lattice
  of size $N.$ $C^M_N$ independent values of $\varphi$ correspond
  to $C^M_N$ values of the wavefunction of $M$ "particles" with coordinates
  $y_i,$ living on this auxiliary lattice (since $\varphi$ is symmetric for $N-M$ fermion indices, these are considered to be vacancies).
  Wavefunction should be symmetric with respect to exchange of two
  "particles" for two-species fermions, and antisymmetric for the
  case of bose-fermi mixture. To preserve the terminology of the two-species fermion solution for the case of bose-fermi
  mixture,  later in the text we will always
  refer to the wavefunction on an auxiliary lattice as to "spin"
  wavefunction, although it has a direct meaning only for
  two-species fermion case.

  First, one can solve the problem for $M=1$\cite{McGuire}. In this case there is no difference between two-species fermions or bose-fermi mixture.
   It can be shown (detailed derivations are available in the appendix of  \cite{Andrei}),
  that in this case  wavefunction in conjugate representation is
  \begin{equation}
  \varphi(M=1)= F(\Lambda,y)=\prod_{j=1}^{y-1}\frac{k_j-\Lambda + ic/2}{{k_{j+1}-\Lambda - ic/2}},
  \label{FLambday}
  \end{equation}
  where  new spectral parameter $\Lambda$ satisfies the following equation:
  \begin{equation}
  \prod_{i=1}^{N}\frac{k_i-\Lambda+ic/2}{k_i-\Lambda-ic/2}=1.
  \end{equation}
 Periodic boundary conditions simplify to
 \begin{equation}
 e^{i k_j L}=\frac{k_j-\Lambda+ic/2}{k_j-\Lambda-ic/2}, j=\{1, ... , N\}
 \end{equation}
In an auxiliary lattice the wavefunction  of one spin-deviate (or
boson) $F(\Lambda,y)$ plays the role similar to one-particle basis
function $e^{i k x}$
 of the original  coordinate Bethe ansatz,
 spectral parameter $\Lambda$ being the analog of the  momentum $k.$

 In the case when $M>1,$  Yang suggested that the solution of eqs.
 (\ref{PBC21})-(\ref{PBC22}) again has the form of Bethe ansatz in the "spin" subspace:
 for $1\leq y_1<y_2< ...< y_M \leq N$
 \begin{equation}
 \varphi=\sum_R A(R) \prod_{i=1}^M F(\Lambda_{R_i}, y_i),
\label{Yangwf}
 \end{equation}
where $\Lambda_1, ... ,\Lambda_M$ is a set of unequal numbers,
$R$ is an arbitrary permutation from $S_M.$ It can be shown
\cite{Yang67,Andrei}, that this ansatz solves
(\ref{PBC21})-(\ref{PBC22}) for two-species fermion system, if
\begin{equation}
\frac{A(R')}{A(R)}=\frac{\Lambda_{R_{i+1}}-\Lambda_{R_{i}}- i c }{\Lambda_{R_{i+1}}-\Lambda_{R_{i}}+ i c },
\label{app}
\end{equation}
similarly to bosonic  relations of Lieb-Liniger model\cite{LL}. Here $R$ and $R'$ are two permutations from $S_M$ such that
 $R_k=R'_k,$ for $k \neq \{i, i+1\}, $ and $R_i=R'_{i+1},R'_i=R_{i+1}.$ The set of $\Lambda, k$ has to satisfy the following set of equations:
\begin{eqnarray}
  -\prod_{i=1}^{N}\frac{k_i-\Lambda_\alpha+ic/2}{k_i-\Lambda_\alpha-ic/2}=\prod_{\beta=1}^{M}\frac{\Lambda_\beta-\Lambda_\alpha+ic}{\Lambda_\beta-\Lambda_\alpha-ic},\alpha=\{1,...,M\},\\
  e^{i k_j L}=\prod_{\beta=1}^{M}\frac{k_j-\Lambda_\beta+ic/2}{k_j-\Lambda_\beta-ic/2}, j=\{1, ... , N\}.
\end{eqnarray}

For the bose-fermi mixture,  $\varphi$ has to be  antisymmetric
for permutations of $y_i$ variables.  This problem has actually
been solved by Sutherland\cite{Sutherland68}, although he was
interested not in bose-fermi mixture, but fermion model with
several species. He has shown, that if one doesn't specify the
symmetry of $\varphi$ for $y_i$ variables  and applies the
generalized ansatz
\begin{equation}
\varphi=\sum_{R} [G,R] \prod_{i=1}^M F(\Lambda_{R_i}, y_{G_i})
\end{equation}
for $1\leq y_{G_1}<y_{G_2}< ...< y_{G_M}\leq N$, then columns of
$M!\times M!$ dimensional matrix $[G,R]$ are related similarly to
(\ref{transfereq}):
\begin{equation}
\xi_{R'}=Y'^{i,i+1}_{R_i,R_{i+1}} \xi_R.
\label{transfereq2}
\end{equation}
$Y'$ operators are defined as
\begin{eqnarray}
Y'^{l,m}_{i,j}=\frac{\kappa_{ij}+ \hat P_{lm}}{1-\kappa_{ij}},
\kappa_{ij}= \frac{ic}{\Lambda_i -\Lambda_j}.
\end{eqnarray}
For two-species fermions in conjugate representation  $ \hat P_{lm}=1,$ and it is
equivalent to (\ref{app}), while for bose-fermi mixture in conjugate representation $\hat
P_{lm}=-1,$ and the answer is much more simple:
\begin{equation}
Y'^{l,m}_{i,j}=-1.
\end{equation}
Therefore, "spin" part of wavefunction is  constructed by total
antisymmetrization of single "spin" wavefunctions, similar to
Slater determinant for fermionic particles:
\begin{equation}
 \varphi=\det(F(\Lambda_i, y_j)).
 \label{bosefermispinwf}
\end{equation}

Periodic boundary conditions for bose-fermi mixture are:
\begin{eqnarray}
  \prod_{i=1}^{N}\frac{k_i-\Lambda_\alpha+ic/2}{k_i-\Lambda_\alpha-ic/2}=1,\alpha=\{1,...,M\},\label{bfequation1}\\
  e^{i k_j L}=\prod_{\beta=1}^{M}\frac{k_j-\Lambda_\beta+ic/2}{k_j-\Lambda_\beta-ic/2}, j=\{1, ... , N\}.
\label{bfequation2}
\end{eqnarray}
 One can prove that all solutions of  (\ref{bfequation1})-(\ref{bfequation2}) are always real, which is a major simplification
for the analysis of both ground and excited states (see Appendix
\ref{appA}).

If one introduces function
\begin{equation}
\theta(k)=-2 \tan^{-1}(k/c),
\end{equation}
the system (\ref{bfequation1})-(\ref{bfequation2}) can be rewritten as
\begin{eqnarray}
k_j L=2 \pi I_j + \sum_{\beta=1}^{M} \theta(2k_j- 2\Lambda_{\beta}),\\
2 \pi I_\alpha=\sum_{j=1}^N \theta(2\Lambda_\alpha-2k_j) .
\label{bfequations2}
\end{eqnarray}
$I_j$ and  $I_\alpha$ are integer or half integer  quantum
numbers (depending on the parity of M and N), which characterize
the state. The ground state corresponds to
\begin{eqnarray}
I_\alpha=(M-1)/2, -(M-3)/2, .... , (M-1)/2\}, \label{ialpha}\\
I_j=\{-(N-1)/2, -(N-3)/2, .... , (N-1)/2\}.\label{ij}
\end{eqnarray}
In the thermodynamic limit, one has to send $M,N,L$ to infinity
proportionally. If one introduces density of  $k$ roots  $\rho(k)$
and  density of $\Lambda$ roots $\sigma(\Lambda),$
 (\ref{bfequation1})-(\ref{bfequation2}) simplifies to two coupled integral equations
\begin{eqnarray}
2 \pi \rho(k)=1 +\int_{-B}^{B}\frac{4c \sigma(\Lambda) d\Lambda }{c^2+4(\Lambda-k)^2} ,\label{bfequations31}\\
2 \pi \sigma(\Lambda)=\int_{-Q}^{Q}\frac{4c \rho(\omega) d\omega }{c^2+4(\Lambda-\omega)^2}.
\label{bfequations32}
\end{eqnarray}
Normalization conditions and energy are given by
\begin{eqnarray}
N/L=\int_{-Q}^{Q}\rho(k) dk, \label{normn} \\
M/L=\int_{-B}^{B}\sigma(\Lambda) d\Lambda, \label{normm}  \\
E/L=\int_{-Q}^{Q} k^2 \rho(k) dk.\label{energy}
\end{eqnarray}

These equations can be solved numerically and the results will be
presented in the next section. Numerical solution of these
equations  allows to investigate the possibility of phase
separation, predicted in \cite{Das}.  Combined with local density
approximation, it can be used to investigate density profiles and
collective oscillation modes in the external fields.

\section{Numerical Solution and Analysis of Instabilities}
\label{numerics}

In this section we will solve the system of equations
(\ref{bfequations31})-(\ref{energy}) numerically, and obtain the
ground state energy as a function of interaction strength and
densities. This solution will be used to analyze the instability
towards demixing\cite{CazalillaHo,Das,jap_numerics}.

\begin{figure}
\psfig{file=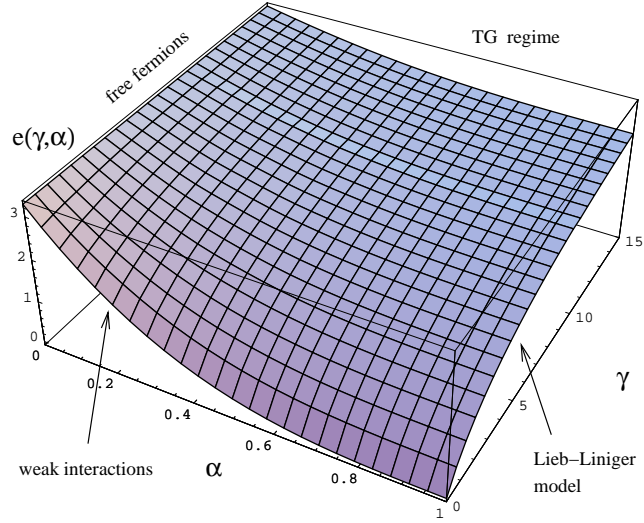} \caption{ \label{egammamn} Energy of
the ground state is given by $E=e(\gamma,\alpha) \hbar^2 N^3 /(2 m
L^2),$ where $\gamma=m g /(\hbar^2 n),$ and $\alpha=M/N$ is the
boson fraction. When $\alpha=0,$ system is purely fermionic, and
the energy doesn't depend on interactions. When $\alpha=1,$ the
system is purely bosonic, and numerically obtained energy
coincides with the result of \cite{LL}. If $\gamma=0,$ bosons and
fermions don't interact, and
$e(\gamma,\alpha)=(\pi^2/3)(1-\alpha)^3.$}
\end{figure}

 Substituting (\ref{bfequations32}) into
(\ref{bfequations31}), and performing analytically integration
over $\Lambda,$  one obtains  an integral equation for function
$\rho(k).$  Similarly to \cite{LL}, it is convenient to redefine
the variables before solving this equation numerically. Let's
introduce the following variables $\lambda, x, y, b$  and a
function $g(x)$ according to
\begin{equation}
c = \lambda Q, \omega=x Q, k=y Q, B = b Q, \rho(Qx)=g(x).
\end{equation}
In new variables, integral equation depends on two parameters $b$ and $\lambda:$
\begin{eqnarray}
2 \pi g(y)=1 +\int_{-1}^{1}\frac{2 \lambda g(x) dx}{2 \pi (\lambda^2+(x-y)^2)}( \tan^{-1}{\frac{2(b-y)}\lambda}+\tan^{-1}{\frac{2(b-x)}\lambda}+  \nonumber \\
 \tan^{-1}{\frac{2(b+y)}\lambda}+ \tan^{-1}{\frac{2(b+x)}\lambda}+
\frac{\lambda}{2(x-y)}\log{\frac{\lambda^2+4(b-x)^2}{\lambda^2+4(b-y)^2}
\frac{\lambda^2+4(b+y)^2}{\lambda^2+4(b+x)^2}}
) .\label{integral}
\end{eqnarray}
In new variables, (\ref{normn})-(\ref{energy}) become
\begin{eqnarray}
\gamma=\frac{c L}{N} = \frac{\lambda}{\int_{-1}^{1}g(x) dx}, \label{gamma} \\
\frac{M}{N}=\frac{\int_{-1}^{1}(\tan^{-1}{\frac{2(b-x)}\lambda}+\tan^{-1}{\frac{2(b+x)}\lambda})g(x) dx}{\pi \int_{-1}^{1}g(x) dx},\label{mn}\\
E=\frac{N^3}{L^2}e(\lambda,b)=\frac{N^3}{L^2}\frac{\int_{-1}^{1}x^2 g(x) dx}{\left(\int_{-1}^{1}g(x) dx\right)^3}\label{energynew}.
\end{eqnarray}
Integral equation (\ref{integral}) can  be solved numerically as a
function of two parameters $b$ and $\lambda,$  applying Simpson
rule for an integral approximation on a grid $x_i=-1+(i-1)/n,
i=\{1, ..., 2n+1\}$. This gives a system of $2n+1$ linear
equations for discrete values $g(x_i),$ which can be solved by
standard methods. Using (\ref{gamma})-(\ref{energynew}), one can
obtain parametrically
 three functions
$\gamma(\lambda,b), M/N=\alpha(\lambda,b), e(\lambda,b).$ After that one can numerically inverse two of
 them $\lambda(\gamma,\alpha)$ and $b(\gamma,\alpha),$ and obtain function $e(\gamma,\alpha).$
Resulting function is shown in fig. \ref{egammamn}. When $\alpha=0,$ system is purely fermionic, and noninteracting.
When $\alpha=1,$ the system is purely bosonic, and numerically obtained energy coincides with the result of \cite{LL}.
If $\gamma=0,$ bosons and fermions don't interact, and $e(\gamma,\alpha)=(\pi^2/3)(1-\alpha)^3.$

An interesting case, where one can analytically find the
dependence of energies on relative densities is Tonks-Girardeau
(TG) regime  of strong interactions, $\gamma \gg 1.$ In
(\ref{integral}) one can neglect the dependence of the kernel on
$x$  and $y,$ and $g(x)$ becomes a constant $g,$ which satisfies
an equation
\begin{equation}
2\pi g=1+ \frac{8g}{\pi \lambda}\left( \tan^{-1}{\frac{2b}{\lambda}}+\frac{2 b \lambda}{\lambda^2+ 4b^2}\right),
\end{equation}
while (\ref{mn}) reads
\begin{equation}
\alpha=\frac2\pi \tan^{-1}{\frac{2b}{\lambda}}.
\end{equation}
After some algebra energy is rewritten as
\begin{equation}
e(\gamma,\alpha)=\frac{\pi^2}3\left(1-\frac4\gamma(\alpha +
\frac{\sin{\pi \alpha}}{\pi})+ \frac{12}{\gamma^2}(\alpha +
\frac{\sin{\pi \alpha}}{\pi})^2\right)+O(\frac{1}{\gamma^3}).
\label{tgenergy}
\end{equation}


Using exact solutions, one can analyze demixing
instabilities\cite{CazalillaHo, Das, jap_numerics} for repulsive
bose-fermi mixtures. In the absence of external potential
bose-fermi mixture is stable, if the  compressibility matrix
\begin{equation}
\left[\begin{array}{cc}
  \frac{\partial\mu_b}{\partial{n_b}} & \frac{\partial\mu_b}{\partial{n_f}} \\
  \frac{\partial\mu_f}{\partial{n_b}} & \frac{\partial\mu_f}{\partial{n_f}}
  \end{array}\right]
\label{compress}
\end{equation}
is positively defined. Here, $n_b$ is the boson density, and $n_f$
is the fermion density. $\mu_b$  and $\mu_f$ are the bose and
fermi chemical potentials, given by
\begin{eqnarray}
\mu_b=\frac{N^2}{L^2}\left( 3e(\gamma,\alpha) -\gamma
\frac{\partial e}{\partial \gamma} +(1-\alpha) \frac{\partial
e}{\partial \alpha}\right),\\ \mu_f=\frac{N^2}{L^2}\left(
3e(\gamma,\alpha) -\gamma \frac{\partial e}{\partial \gamma}
-\alpha \frac{\partial e}{\partial \alpha}\right).
\end{eqnarray}

The fact that the matrix (\ref{compress}) is positively defined
can be checked numerically for any value of $\alpha$ and $\gamma$,
and proves that bose-fermi mixture with the same bose-fermi and
bose-bose interactions is stable with respect to demixing for any
values  of bose and fermi densities. We note, that the absence of
demixing for one particular value of the density has been checked
in the original article by Lai and Yang\cite{Lai}. Although an
exact solution is available only under conditions
(\ref{intercond}), small deviations from these should not
dramatically change the energy $e(\gamma,\alpha).$ Therefore, we
expect the 1D mixtures to remain stable to demixing in the
vicinity of the integrable line (\ref{intercond}) for any
interaction strength. Recently this has been checked numerically in
Quantum Monte Carlo studies for a systems of up to $14$ atoms\cite{jap_numerics}.

Note, that prediction of \cite{Das} about demixing  at sufficiently
strong interactions in this case is incorrect, since it is based
on the mean field approximation. Indeed, the demixing condition
there reads
\begin{equation}
n_f\leq \frac{m_f g_{bf}^2}{g_{bb}\hbar^2 \pi^2 }.
\end{equation}
For $g_{bf}=g_{bb}$ and $n_b=n_f$ it is equivalent to
\begin{equation}
\gamma=\frac{m g}{\hbar^2 n}\geq \frac{\pi^2}2\approx 4.9.
\end{equation}
Clearly, this condition is incompatible with mean-field
approximation, which is valid for  $\gamma \lesssim 1.$


For weakly interacting case one can use mean field approximation
to calculate energy  and chemical potentials\cite{Das, Batchelor}
:
\begin{eqnarray}
E=L\left[\frac{g}{2}n_b^2+g n_b n_f+\frac{\hbar^2 \pi^2}{2m}
\frac{n_f^3}{3} \right], \nonumber \\ \mu_b=g(n_b+n_f), \; \mu_f=g
n_b+\frac{\hbar^2 \pi^2}{2m} n_f^2. \label{Meanfield}
\end{eqnarray}
For the strong interactions, up to corrections of order
$1/\gamma^3$,
\begin{eqnarray}
E=L\left[\frac{\hbar^2 \pi^2}{2m} \frac{(n_f+n_b)^3}{3}
\right]\left(1-\frac4\gamma(\alpha + \frac{\sin{\pi
\alpha}}{\pi})+ \frac{12}{\gamma^2}(\alpha + \frac{\sin{\pi
\alpha}}{\pi})^2\right), \label{largegammaenergy}
\\ \mu_f= \frac{\hbar^2 \pi^2}{2m}
(n_f+n_b)^2 (1+\frac1{3\gamma}\left(-16(\alpha + \frac{\sin{\pi
\alpha}}{\pi})+4 \alpha(1 + \cos{\pi \alpha})\right)+ \nonumber \\
+\frac{1}{\gamma^2}(\alpha + \frac{\sin{\pi
\alpha}}{\pi})\left(20(\alpha + \frac{\sin{\pi
\alpha}}{\pi})-8\alpha(1 + \cos{\pi \alpha})\right)),\\
\mu_b=\frac{\hbar^2 \pi^2}{2m} (n_f+n_b)^2
(1+\frac1{3\gamma}\left(-16(\alpha + \frac{\sin{\pi
\alpha}}{\pi})+4 (\alpha-1)(1 + \cos{\pi \alpha})\right)+
\nonumber
\\ +\frac{1}{\gamma^2}(\alpha + \frac{\sin{\pi
\alpha}}{\pi})\left(20(\alpha + \frac{\sin{\pi
\alpha}}{\pi})+8(1-\alpha)(1 + \cos{\pi \alpha})\right)).
\end{eqnarray}

\section{Local density approximation and collective modes}
\label{LDAsection}
So far our arguments have been limited to the case of periodic
boundary conditions without external confinement. This is the
situation, when the many-body interacting model
(\ref{initialhamiltoniansc}) is exactly solvable in the
mathematical sense. If one adds an external harmonic potential,
model is not solvable any more. However, if external potential
varies slowly enough (precise conditions for the case of bose gas
have been formulated in \cite{ShlyapnikovLDA}), one can safely use
local density approximation (LDA) to analyze the density profiles
and collective modes in a harmonic trap. In the local density
approximation, one assumes that  in slowly
varying external harmonic trap chemical potential changes according to
\begin{equation}
\mu^0_b(x)+ \frac{m \omega^2_b x^2}2=\mu^0_b(0), \; \mu^0_f(x)+
\frac{m \omega^2_f x^2}2=\mu^0_f(0). \label{LDA}
\end{equation}

\begin{figure}
\psfig{file=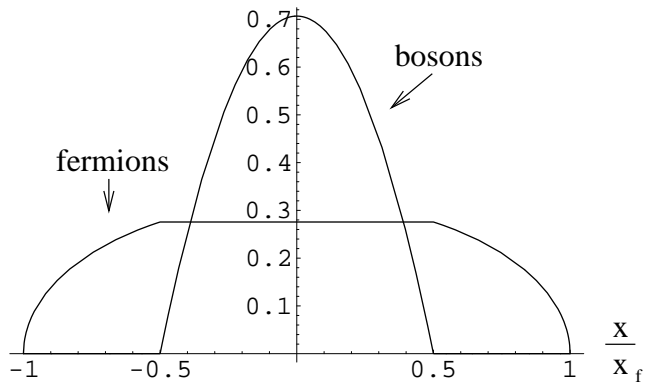} \caption{ \label{meanfigure}  Densities
of bose and fermi gases in weakly interacting regime at zero temperature. Lieb-Liniger
parameter in the center of a trap is $\gamma_0=0.18,$ overall
number of bosons equals number of fermions. Total density in the center of a trap is taken to be $1$.}
\end{figure}

\begin{figure}
\psfig{file=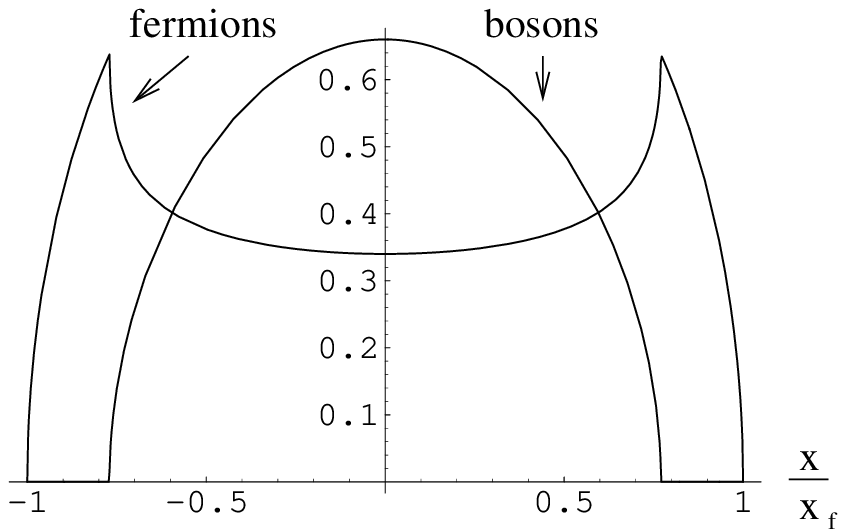} \caption{ \label{lgdp} Densities of bose and
fermi gases in strongly interacting regime at zero temperature. Lieb-Liniger
parameter in the center of a trap is $\gamma_0\gg1,$ overall
number of bosons equals number of fermions. Total density in the center of the trap is taken to be $1$.}
\end{figure}

 Let us
consider the case when external harmonic confining potential
oscillator frequencies are the same for bosons and fermions. We
note, however, that one can  also analyze the case when $\omega_b
\neq \omega_f$ in a similar way. We consider
\begin{equation}
\omega_b= \omega_f=\omega_0, \label{equalomega}
\end{equation}
since in this case distribution of
the relative boson and fermion densities is controlled only by
interactions, and not by external potential, since external potential couples only
to total density. Eqs. (\ref{LDA}) for
$\omega_b=\omega_f=\omega_0$ imply that densities of bosons and
fermions  in the region where bosons and fermions coexist are
governed by
\begin{eqnarray}
\mu^0_f(x)+ \frac{m \omega^2_0 x^2}2=\mu^0_f(0),\;
\mu^0_b(x)-\mu^0_f(x)=(n_f+n_b)^2 \frac{\partial e}{\partial
\alpha}=\mu^0_b(0)-\mu^0_f(0). \label{LDA2}
\end{eqnarray}
One can show, that these equations cannot be simultaneously
satisfied for the whole cloud, and the mixture phase separates in
an external potential given by (\ref{equalomega}). For both strong
and weak interactions bosons and fermions coexist in the central
part, but the outer sections consist of Fermi gas only. In the
weakly interacting limit, this can be interpreted as an effect of
the Fermi pressure\cite{fermipressure}: while bosons can condense
to the center of the trap, Pauli principle pushes fermions apart.
As interactions get stronger, the relative distribution of bosons
and fermions changes, and Figs. \ref{meanfigure} and \ref{lgdp}
contrast the limits of strong and weak interactions. For strong
interactions, the fermi density shows strong non-monotonous
behavior.

When interactions are small Eqs. (\ref{Meanfield}) and
(\ref{LDA2}) imply that in the region of coexistence densities are
given by
\begin{eqnarray}
n^0_b(x)=n^0_b(0)(1-\frac{x^2}{x_b^2}),\; n^0_f(x)=n^0_f(0), \;
\mbox{for} \; x^2<x_b^2. \label{mfdensitiesinside}
\end{eqnarray}
Outside of the region of coexistence, density of fermions decays
as the square root of inverse parabola:
\begin{equation}
n^0_b(x)=0,\;
n^0_f(x)=\frac{n^0_f(0)}{\sqrt{1-\frac{x_b^2}{x_f^2}}}\sqrt{1-\frac{x^2}{x_f^2}}
, \; \mbox{for} \; x_b^2<x^2<x_f^2 . \label{mfdensitiesoutside}
\end{equation}
 Parameters $x_f$ and $x_b$ are given by
\begin{eqnarray}
x_b^2=\frac{2 g n^0_b(0)}{m \omega_0^2 }, \; x_f^2=
x_b^2+\frac{(\hbar \pi n^0_f(0))^2}{( m\omega_0)^2}.
\end{eqnarray}

A typical graph of density distribution for weakly interacting case is shown is shown in Fig. \ref{meanfigure}.

If effective $\gamma_0$ is much bigger than $1$ in the center of
a harmonic trap, the total density $n^0(x)$ follows Tonks-Girardeau density profile
\begin{equation}
n^0(x)=n^0(0) \sqrt{1-\frac{x^2}{x_f^2}}.
\label{tgtotaldp}
\end{equation}
From Eqs. (\ref{tgenergy}) and (\ref{LDA2}) distribution of $\alpha(x)$ is controlled by  the following equation:
\begin{equation}
n^0(x)^3(1+\cos{(\pi \alpha(x))})=n^0(0)^3(1+\cos{(\pi
\alpha(0))}).
\end{equation}
Since $1+\cos{(\pi \alpha(x))}$ is bound and $n^0(x)$ goes to $0$
near the edges of the cloud, this equation can't be satisfied for
all $x^2< x_f^2,$ which means that only fermions will be present
at the edges of the cloud, similarly to weakly interacting regime.
Density distribution for equal number of bosons and fermions is
shown in Fig. \ref{lgdp}. The form of the profile is universal, as long as $\gamma_0\gg1$ and the temperature is zero. Evolution of this profile for nonzero temperatures is shown in figure \ref{nztdensity}.

Recent experiments\cite{Moritz} demonstrated that collective
oscillations of 1D gas provide useful information about
interactions in the system. Here we will numerically investigate
collective modes of the system, by solving hydrodynamic equations
of motion. These equations have to be solved with proper boundary
conditions at the edge of  the bosonic and fermionic clouds.
Within the region of coexistence of bosons and fermions, such
oscillations can be described by four hydrodynamic
equations\cite{Menotti}
\begin{eqnarray}
\frac{\partial}{\partial t}\delta n_b+\frac{\partial}{\partial
x}(n_b v_b)=0,\label{dnbdt}\\ m\frac{\partial}{\partial t}\delta
v_b+\frac{\partial}{\partial x}(\mu_b+V_{ext, b}+\frac12 m
v_b^2)=0,\label{dvbdt}\\ \frac{\partial}{\partial t}\delta
n_f+\frac{\partial}{\partial x}(n_f v_f)=0,\label{dnfdt}\\
m\frac{\partial}{\partial t}\delta v_f+\frac{\partial}{\partial
x}(\mu_f+V_{ext, f}+\frac12 m v_f^2)=0\label{dvfdt}.
\end{eqnarray}

In certain cases, analytical solutions of hydrodynamic equations
are available\cite{Stringari, Menotti} and provide the frequencies
of collective modes. When an analytic solution is not available,
the "sum rule" approach has been used\cite{Stringari,
Menotti,Astrakharchik,bfsumrule}  to obtain an upper bound for the
frequencies of collective excitations. The disadvantage of the
latter approach is an ambiguity in the choice of multipole
operator which excites a particular mode, especially for
multicomponent systems\cite{bfsumrule}. Here we develop an
efficient numerical procedure for solving hydrodynamical equations
in 1D, which doesn't involve additional "sum rule" approximation.

While looking at  low amplitude oscillations, it is sufficient to
substitute
\begin{eqnarray}
n_b(x,t)=n_b^0(x)+\delta n_b(x) e^{i \omega t },\label{dntnb}\\
n_b(x,t) v_b(x,t)=n^0_b(x)\delta v_b(x) e^{i \omega t },\\
n_f(x,t)=n_f^0(x)+\delta n_f(x) e^{i \omega t }, \label{dntnf}\\
n_f(x,t) v_f(x,t)=n^0_f(x,t)\delta v_f(x) e^{i \omega t },\\
\mu_b+V_{ext, b}+\frac12 m v_b^2=const1+ \delta \mu_b(x) e^{i
\omega t }=  const1+(\delta n_b(x)\frac{\partial \mu_b}{\partial
n_b}+\delta n_f(x)\frac{\partial \mu_b}{\partial n_b})e^{i \omega
t },\label{dvdtb}\\ \mu_f+V_{ext, f}+\frac12 m v_f^2=const2+
\delta \mu_f(x) e^{i \omega t }=  const2+(\delta
n_b(x)\frac{\partial \mu_f}{\partial n_b}+\delta
n_f(x)\frac{\partial \mu_f}{\partial n_f})e^{i \omega t }.
\label{dvdtf}\\
\end{eqnarray}
Here, $n_b^0(x)$ and $n_f^0(x)$ are densities obtained within
local density approximation. Linearized system of hydrodynamic
equations can be written as:
\begin{eqnarray}
- m \omega ^2 \left[\begin{array}{c}
\delta n_b(x) \\
\delta n_f(x)
\end{array}\right] =
\nabla\left(\left[ \begin{array}{cc}
  n^0_b(x) & 0 \\
  0 & n^0_f(x)
  \end{array}\right]
  \nabla \left(
\left[\begin{array}{cc}
  \frac{\partial\mu_b}{\partial{n_b}} & \frac{\partial\mu_b}{\partial{n_f}} \\
  \frac{\partial\mu_f}{\partial{n_b}} & \frac{\partial\mu_f}{\partial{n_f}}
  \end{array}\right]\left[\begin{array}{c}
\delta n_b(x) \\
\delta n_f(x)
\end{array}\right]\right)\right).
\end{eqnarray}
For numerical solutions and boundary conditions  it is more convenient to work with independent functions $\delta \mu_b(x), \delta \mu_f(x).$
System of equations becomes
\begin{eqnarray}
- m \omega ^2 \left[\begin{array}{c}
\delta \mu_b(x) \\
\delta \mu_f(x)
\end{array}\right] =
\left[\begin{array}{cc}
  \frac{\partial\mu_b}{\partial{n_b}} & \frac{\partial\mu_b}{\partial{n_f}} \\
  \frac{\partial\mu_f}{\partial{n_b}} & \frac{\partial\mu_f}{\partial{n_f}}
  \end{array}\right]\nabla\left(\left[ \begin{array}{cc}
  n^0_b(x) & 0 \\
  0 & n^0_f(x)
  \end{array}\right]
  \nabla
\left[\begin{array}{c}
\delta \mu_b(x) \\
\delta \mu_f(x)
\end{array}\right]\right).
\label{eqmuinside}
\end{eqnarray}
Outside of the region of coexistence of bosons and fermions, $\delta \mu^{out}_f$ satisfies the following equation:
\begin{equation}
- m \omega ^2 \delta \mu^{out}_f=\frac{\partial \mu_f^{out}}{\partial{n_f}}\nabla\left[n^{out}_f(x)\nabla \delta \mu^{out}_f \right].
\label{eqout}
\end{equation}

All modes can be classified by their parity with respect to
$x\rightarrow-x$ substitution, and will be investigated by
parity-dependent numerical procedure. We will consider equations
only in the positive half of the cloud. For even modes, one may
require two additional conditions:
\begin{eqnarray}
\nabla \delta \mu_f(x=0)=0, \, \nabla \delta
\mu_b(x=0)=0.\label{evenbc}
\end{eqnarray}
For odd modes, analogous  conditions are
\begin{eqnarray}
\delta \mu_f(x=0)=0, \,  \delta \mu_b(x=0)=0.\label{oddbc}
\end{eqnarray}

Boundary conditions for fermions at the edge of the bosonic cloud,
$x_b,$ correspond to the continuity of $v_f$ and $\delta \mu_f.$
Continuity of the velocity can be obtained by integrating
continuity equation  (\ref{dnfdt}) in the
 vicinity of $x_b.$ From Eq. (\ref{dvfdt})  it is equivalent to
\begin{equation}
\nabla \delta \mu^{out}_f (x=x_b+0)=\nabla \delta \mu_f(x=x_b-0).
\label{boundc1}
\end{equation}

The second condition can be obtained by integrating (\ref{dvfdt})
in the vicinity of $x_b:$
\begin{equation}
\delta \mu^{out}_f (x=x_b+0)= \delta \mu_f(x=x_b-0).
\label{boundc2}
\end{equation}

One may see, that these conditions {\it do not} imply that $\delta
n^{out}_f (x=x_b+0)=\delta n_f (x=x_b-0).$ This can be easily
illustrated by the dipole mode, where $\delta v_f(x)=\delta
v_b(x)=const, \delta n_f = \nabla n_f^0(x), $ which is clearly
discontinuous for profiles shown in Figs. \ref{meanfigure} and
\ref{lgdp}.

Two additional conditions come from the absence of the
bosonic(fermionic) flow at $x_b (x_f):$
\begin{eqnarray}
n^0_b(x) v_b(x)\vert_{x\rightarrow x_b-0}=0,\label{boseflow}\\
n^0_f(x) v_f(x)\vert_{x\rightarrow x_f-0}=0. \label{fermiflow}
\end{eqnarray}

Outside of the region of coexistence, the chemical potential and
density of fermions are given by  $\mu^{out}_f\sim (n^{out}_f)^2,
n^{out}_f \sim\sqrt{1-(x/x_f)^2},$ where $x_f$ is the fermionic
cloud size. In dimensionless variables $u=x/x_f,$ eq. (\ref{eqout})
can be written as 
\begin{equation}
- \frac{\omega ^2}{\omega_0^2} \delta \mu^{out}_f=(1-u^2)\frac{\partial^2 \delta \mu_f^{out}}{\partial{u}^2} - u \frac{\partial \delta \mu_f^{out}}{\partial{u}}.
\end{equation}
For this equation, there exists a general nonzero
solution  which satisfies (\ref{fermiflow}):
\begin{equation}
\delta \mu_f^{out}= \cos(\frac{\omega}{\omega_0}\arccos{\frac{x}{x_f}}).
\end{equation}

Substituting this into (\ref{boundc1})-(\ref{boundc2}), one has to
solve eigenmode equations numerically for $x<x_b,$ with five
boundary conditions
(\ref{boundc1}),(\ref{boundc2}),(\ref{boseflow}) and
(\ref{evenbc}) or (\ref{oddbc}) depending on the parity. These
boundary conditions are compatible, only if $\omega$ is an
eigenfrequency. Using four of these boundary conditions, the
system of two second order differential equations can be solved
numerically for any $\omega.$ To find a numerical solution we
choose to leave out condition (\ref{boseflow}), and check later if
it is satisfied to identify the eigenfrequencies.

The most precise way to check (\ref{boseflow}) numerically is
based on equations of motion. For even modes, $v_b(0)=0,$ and
integrating (\ref{dnbdt}) from $0$ till $x_b,$ one obtains
\begin{equation}
\int_0^{x_b}\delta n_b(x) dx=-\frac1{i\omega}(n_b(x=x_b) v_b(x=x_b)-n_b(x=0)v_b(x=0)])=0.\label{evenintegral}
\end{equation}
For odd modes, from eq. (\ref{dvbdt}) $v_b(0)=i \nabla \delta
\mu_b(x=0)/(m\omega),$ and integrating (\ref{dnbdt}) from $0$ till
$x_b,$ one obtains
\begin{equation}
\int_0^{x_b}\delta n_b(x) dx=-\frac1{i\omega}(n_b(x=x_b)v_b(x=x_b)-n_b(x=0)v_b(x=0)])=\frac{n_b(x=0)\nabla \delta \mu_b(x=0)}{m\omega^2}.\label{oddintegral}
\label{normeven}
\end{equation}
When a numerical solution for $\delta \mu_b(x),\delta \mu_f(x)$ is available, conditions (\ref{evenintegral}) or (\ref{oddintegral}) can be checked
numerically using
\begin{equation}
\delta n_b(x)=\frac{\frac{\partial \mu_f }{\partial n_f}\delta \mu_b(x) - \frac{\partial \mu_f }{\partial n_b}\delta \mu_f(x)}{\frac{\partial \mu_f }{\partial n_f}\frac{\partial \mu_b }{\partial n_b}-\frac{\partial \mu_f }{\partial n_b}\frac{\partial \mu_b }{\partial n_f}}.
\label{normodd}
\end{equation}

\begin{figure}
\psfig{file=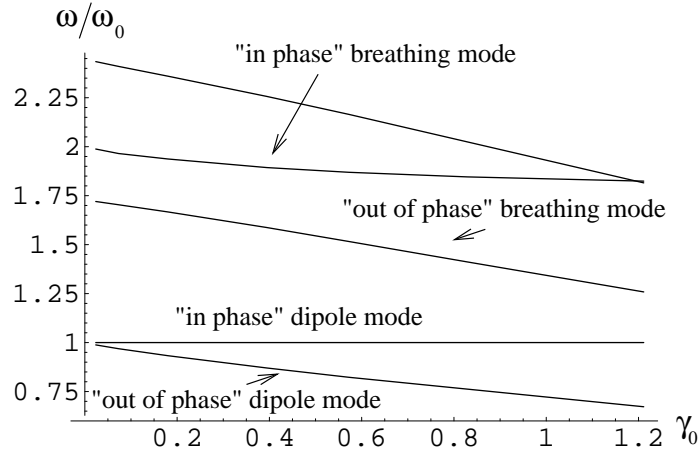} \caption{ \label{mfmodes}
 Frequencies of collective excitations  in mean field regime
  versus Lieb-Liniger parameter in the center of  a trap $\gamma_0.$
  Total number of bosons equals the number of fermions. Even in mean field regime frequency of "out of phase" oscillations gets smaller as interactions get stronger.   }
\end{figure}

\begin{figure}
\psfig{file=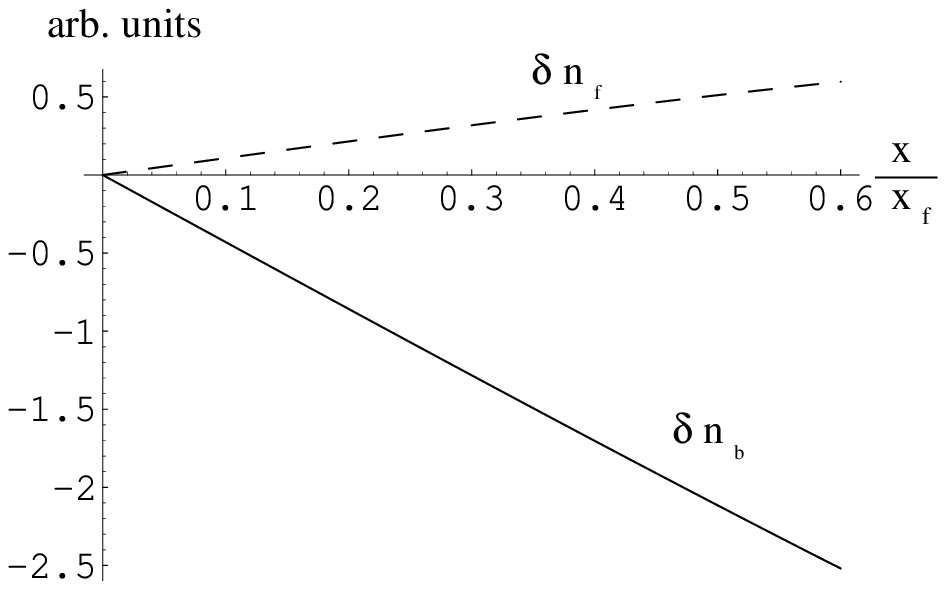} \caption{ \label{n1outofphase}
Fermi and bose density fluctuations of "out of phase" dipole mode $(\omega/\omega_0)^2=0.757$ for $\gamma_0=0.394, x_b/x_f=0.6.$ Total number of bosons equals the number of fermions. Outside of the region of coexistence of bose and fermi clouds, $\delta n_f(x)\sim \frac1{\sqrt{1-(x/x_f)^2}}\cos(\frac{\omega}{\omega_0}\arccos{\frac{x}{x_f}})$}
\end{figure}

\begin{figure}
\psfig{file=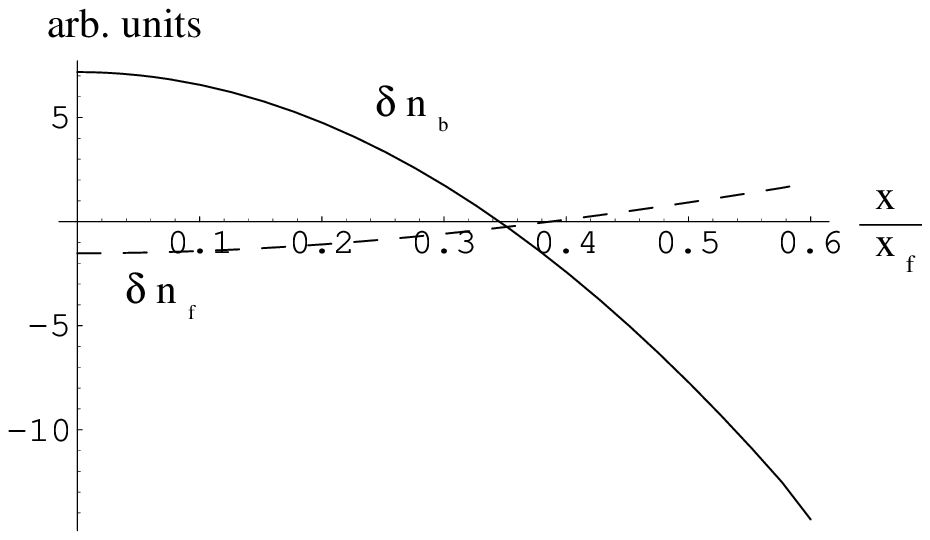} \caption{ \label{n2outofphase}
Fermi and bose density fluctuations of "out of phase" breathing mode $(\omega/\omega_0)^2=2.51$ for $\gamma_0=0.394, x_b/x_f=0.6.$ Total number of bosons equals the number of fermions. Outside of the region of coexistence of bose and fermi clouds, $\delta n_f(x)\sim \frac1{\sqrt{1-(x/x_f)^2}}\cos(\frac{\omega}{\omega_0}\arccos{\frac{x}{x_f}})$}
\end{figure}

\begin{figure}
\psfig{file=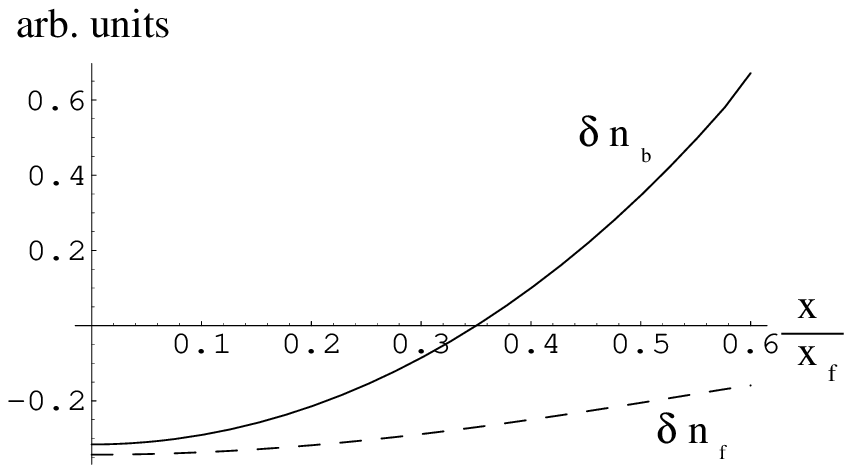} \caption{ \label{n2inphase}
Fermi and bose density fluctuations of "in phase" breathing mode $(\omega/\omega_0)^2=3.585$ for $\gamma_0=0.394, x_b/x_f=0.6.$ Total number of bosons equals the number of fermions. Outside of the region of coexistence of bose and fermi clouds, $\delta n_f(x)\sim \frac1{\sqrt{1-(x/x_f)^2}}\cos(\frac{\omega}{\omega_0}\arccos{\frac{x}{x_f}})$}
\end{figure}

First we apply this numerical procedure for weakly-interacting 
regime, and the frequencies of collective modes are shown in  fig. \ref{mfmodes}.
When $\gamma_0\rightarrow 0,$ bose and fermi clouds do not interact,
and collective modes coincide with purely bosonic or fermionic modes,
with frequencies\cite{Menotti} $\omega^f=n \omega_0$ and $\omega^b=\omega_0 \sqrt{n(n+1)/2}.$ Modes which correspond to $\omega/\omega_0=1, \sqrt{3}, 2, \sqrt{6}$ are shown in
fig. \ref{mfmodes}. As interactions get stronger, bose and fermi clouds get coupled,
and all the modes except for  Kohn dipole mode change their frequency.
For Kohn dipole mode, bose and fermi density fluctuations are given by 
$\delta n_f = \nabla n_f^0(x),\delta n_b = \nabla n_b^0(x).$
In Figs \ref{n1outofphase}-\ref{n2inphase} we show density fluctuations
for three other modes in the region of coexistence for a particular choice of parameters 
$\gamma_0=0.394, x_b/x_f=0.6$ and equal total number of bosons and fermions.
Modes for which the frequency goes down due to coupling  between bose and fermi clouds correspond to the collective excitations with opposite signs in density fluctuations of bose and fermi clouds. In TG regime these modes continuously transform into "out of phase"
low-lying modes which do not change the total density.
At weak interactions lowest mode is an "out of phase" dipole excitation,  after that comes "in phase" Kohn dipole mode (center of mass oscillation), "out of phase" even
mode, "in phase" even mode, second "out of phase" odd  mode.

 \begin{figure}
\psfig{file=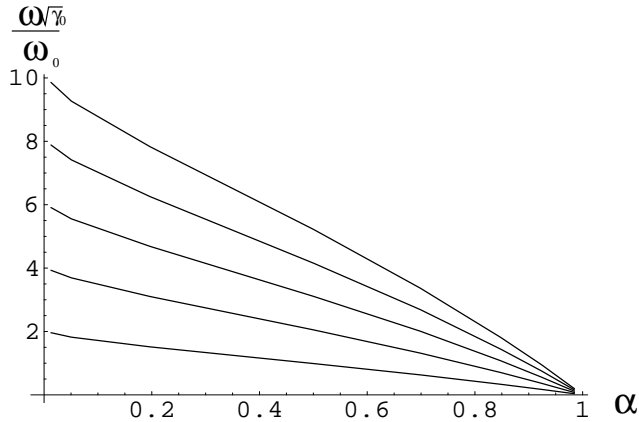} \caption{ \label{tgmodes} Dependence of
the frequency of lowest lying  "out of phase" modes for $\gamma_0 \gg 1$
on overall boson fraction $\alpha,$ where $\gamma_0$ is the Lieb-Lininger
parameter in the center of  a trap. Characteristic scale of "out of phase" oscillations in strongly interacting regime is $\omega_0/\sqrt{\gamma_0} \ll \omega_0.$   Total density "in phase" modes $\omega_n=n \omega_0$ have much higher frequency for $\gamma_0 \gg 1$ and are not shown here. }
\end{figure}

Let's consider Tonks-Girardeau regime, when energy is well
approximated by (\ref{tgenergy}). Since dependence of the energy
on relative boson fraction $\alpha(x)$ is $1/\gamma$ times smaller
than dependence on the total density, the energetic penalty for
changing relative density of bosons and fermions is small. Thus
there should be low-lying modes, which correspond to  an
oscillation of the relative density between bosons and fermions,
while total density is kept fixed up to $1/\gamma$ corrections. In
addition to  these low-lying "out of phase" oscillations of bose
and fermi clouds, there will be "in phase" density modes, which
correspond to oscillations of the total density. Since up to
$1/\gamma$ corrections dependence of the energy on total density
in TG regime is the same as for free noninteracting fermions,
energy of these excitations is given by \cite{Menotti} $\omega=n
\omega_0,$ up to small corrections of the order of $1/\gamma.$

When $\gamma \rightarrow \infty,$ relative compressibility  goes
to zero as $1/\gamma,$ so from eq. (\ref{eqmuinside}) energy of
low-lying modes goes to zero as $1/\sqrt{\gamma_0},$ where $\gamma_0$ is a Lieb-Liniger parameter in the center of  a trap. Performing a
numerical procedure outlined above, one can obtain the dependence
of the frequencies of low-lying "out of phase" modes on relative
density of bosons and fermions. Results of these calculations are
shown in fig. \ref{tgmodes}, and are parameterized by the  overall
boson fraction and $\gamma_0.$  It turns out that the lowest lying mode is
odd, and after that the parity of collective excitations
alternates  signs.  For "out of phase" modes signs of density
fluctuations and velocities of boson and fermion clouds are opposite.
One can easily understand, why does the energy grow, as the boson
fraction is decreased: the size of the bose cloud shrinks, and the
"wavevector" of the  corresponding excitation increases, leading
to an increase of the frequency. One should note that for very small
overall boson fraction  $\gamma_0 \gg 1$ is not enough to separate energy scales for "out of phase" and "in phase" oscillations, and also conditions for applicability of LDA
become more stringent.

\section{Zero-temperature correlation Functions in Tonks-Girardeau regime}
\label{zerotcorrfunc}
 Calculation of the collective modes in the previous section
relies only on the dependence of the energy $e(\gamma, \alpha)$ on
the densities of bosons and fermions. Collective modes can be used
in experiments \cite{Moritz,3Dmodes} to check to some extent
quantitatively the equation of the state of the
system\cite{3dastr}. However, only some part of the information
about the ground state properties is encoded in the energy:
indeed, the energy and collective modes of the strongly
interacting Lieb-Liniger gas are the same as for the free
fermions\cite{girardeau,LL}, while the correlation functions are
dramatically different\cite{Lenard}. Single particle correlation
functions can be measured experimentally using Bragg
spectroscopy\cite{Bragg} or time of flight
measurements\cite{Paredes}. Generally, it is much harder to
calculate the correlation functions compared to the energy from
Bethe ansatz solution. Most of the progress in this direction has
been achieved for the case of strong interactions\cite{KBI}.
Recently, there have been some reports \cite{Carmelo}, where pseudofermionization
method has been used to calculate correlation functions for
spin-$\frac12$ fermion Hubbard model for the intermediate
interaction strengths. In this section, we will analyze the
correlation functions in the regime of strong interactions, using
the factorization of orbital and "spin" degrees of freedom
similarly to the case of spin-$\frac12$ fermions\cite{Woynarovich,OgataShiba}.
Our calculations in this section are performed for
the periodic boundary conditions, when the many body problem is
strictly solvable in the mathematical sense. We will obtain a
representation of correlation functions through the determinants
of some matrices, with the size of these matrices scaling linearly
with the number of the particles. These determinants can be easily
evaluated numerically, and provide a straightforward way to study
correlation functions quantitatively at all distance scales. This
determinant representation can be generalized to nonzero
temperatures, and results  of this generalization will be
presented in the next section.

\subsection{Factorization of "spin" and orbital degrees of freedom}
The regime of strong interactions can be investigated in by
neglecting $k_i$ compared to $\Lambda_\alpha, c $ in
(\ref{bfequation1})-(\ref{bfequation2}). Simplified system for
spectral parameters is
\begin{eqnarray}
 \left(\frac{-\Lambda_\alpha+ic/2}{-\Lambda_\alpha-ic/2}\right)^N=1,\alpha=\{1,...,M\},\label{spectral1}\\
  e^{i k_j L}=\prod_{\beta=1}^{M}\frac{-\Lambda_\beta+ic/2}{-\Lambda_\beta-ic/2}, j=\{1, ... , N\}.
\label{spectral2}
\end{eqnarray}
We see that "spin" part is decoupled from orbital degrees in the
Bethe equations. Equation (\ref{spectral1}) for ground state
"spin" rapidities can be resolved as
\begin{equation}
\frac{-\Lambda_\alpha+ic/2}{-\Lambda_\alpha-ic/2}=e^{i 2\pi
\kappa_\alpha/N},  \alpha=\{1,...,M\}, \label{spinrap}
\end{equation}
where $\kappa_\alpha$ is a set of  integer "spin" wave vectors.
Since the details of calculations depend on the parity of $M$ and
$N$, from now on we will assume that $N $ is even, and
$M$ is odd. Ground state corresponds to $\Lambda_\alpha$
occupying "Fermi sea" $(-\Lambda, \Lambda),$ so from
(\ref{spinrap}) ground state "spin" wave vectors are
\begin{equation}
\kappa_i=\{ -(M-1)/2+N/2, ... , N/2 , ..., (M-1)/2 + N/2\}.
\label{kappaset}
\end{equation}
This choice of "spin" wave vectors will be justified later, in
section \ref{nztcorrelations}. From equation (\ref{spectral2}) it
follows that ground state orbital wave vectors are
\begin{equation}
k_i=\{ -\pi(N-1)/L, ... , -\pi/L ,\pi/L, ... , \pi(N-1)/L\}.
\label{kset}
\end{equation}

Eq. (\ref{FLambday}) for $F(\Lambda,y)$ simplifies to
\begin{equation}
F(\Lambda_\alpha, y_i)=\left(\frac{-\Lambda_\alpha+i
c/2}{\Lambda_\alpha+i c/2}\right)^{y_i-1}=e^{i \frac{2\pi
}{N}\kappa_\alpha(y_i-1)},
\end{equation}
and "spin" wavefunction  (\ref{bosefermispinwf}) can be
represented as a Slater determinant of $M$ single particle plane
waves in "spin" space:
\begin{equation}
\varphi= e^{-i \frac{2\pi }{N}\sum
\kappa_\alpha}\det[e^{i\frac{2\pi}{N} \kappa_i y_j}].
\label{varphiasadet}
\end{equation}
Orbital part of the wavefunction also simplifies  into a Slater
determinant, since all Yang matrices $Y^{a,b}_{i,j}$  in
(\ref{Yangm}) are equal to $-1$.

Ground state is written as a product of two Slater determinants,
describing orbital and "spin" degrees of freedom:
\begin{equation}
\Psi(x_1,...,x_N)\sim \det[e^{i k_i x_j}] \det[e^{i\frac{2\pi}{N}
\kappa_i y_j}]. \label{tgwf}
\end{equation}
Here $x_1, ..., x_M$ are coordinates of bosons, $x_{M+1},..., x_N$
are coordinates of fermions, and $y_i$ is the order in which the
particle $x_i$ appears, if the set $x_1, ..., x_N$ is ordered. In
other words, if
\begin{eqnarray}
0\leq x_{Q_1}\leq x_{Q_2}\leq....\leq x_{Q_N}\leq L,
\label{simplex}
\\
\mbox{then}\ \; \{y_1,...,y_N\}=\{Q^{-1}(1), ... ,Q^{-1}(N)\}.
\end{eqnarray}
  First determinant depends on positions of both bosons and fermions,
while the second determinant depends only on relative positions of
bosons $y_1, ..., y_M.$ Normalization prefactor will be determined
later to give a correct value of the density. One can confirm that
symmetry properties of wavefunction are as required: transposition
of two fermions affects only first determinant, therefore
wavefunction acquires $-1$ sign. Transposition of two bosons
changes signs of both first and second determinants, so
wavefunction doesn't change.

Similar factorization of wavefunction into spin and orbital
degrees of freedom has been observed in \cite{OgataShiba}  for one
dimensional spin-$\frac12$ Hubbard model. In that case, spin
wavefunction is a ground state of spin-$\frac12$ antiferromagnetic
Heisenberg model, and is much more complicated  compared to
(\ref{varphiasadet}).

 It might seem that "spin" degrees are now
independent of orbital degrees,  but this is not true, since  it
is the relative position of orbital degrees which determines
"spin" coordinates. If one wants to calculate, say, bose-bose
correlation function, one has to fix position of $x_1$ and $x'_1$
and integrate $\Psi(x_1,x_2,..,x_N)\Psi^{\dagger}(x'_1,
x_2...,x_N)$ over $x_2, ..., x_N.$ However, there are $C_N^M$
inequivalent spin distributions, and integration in each subspace
(\ref{simplex}) has to be performed separately. For spin-$\frac12$
fermions on a lattice in \cite{OgataShiba} this  integration
becomes a summation, and it has been done numerically for up to
$32$ cites. This summation requires computational resources which
scale as an  exponential of the  number of particles. Here we will
report a method to perform integrations for a polynomial time,
which will allow to go for larger system sizes (easily up to $100$
on a desktop PC) and study correlation functions much more
accurately.

\subsection{Bose-Bose correlation function}

Let's describe a procedure to calculate bose-bose correlation
functions of the model. First, we will use translational symmetry
of the model to fix the positions of the first particle at points
$x_1=0, x'_1 =\xi.$ Instead of writing wavefunction as a function
of positions of  $M$ bosons and $N-M$ fermions,  let's introduce a
set of $N$ ordered variables
\begin{equation}
Z=\{0\leq z_1 \leq z_2 \leq ...  \leq z_N \leq L\},
\label{zhatset}
\end{equation}
which describe positions of the atoms, without specification of
bosonic or fermionic nature of the particle. If any two particles
exchange their positions, they are described by the same set
(\ref{zhatset}). In addition to (\ref{zhatset}) one has to
introduce a permutation $ \hat y$ which specifies positions of
bosons: $y_1, ... ,y_M$ are boson positions, and $y_{M+1}, ...
,y_N$ are fermion positions in an auxiliary lattice:
$z_{y_i}=x_i.$ In this new parameterization normalized
wavefunction is(normalization will be derived later in this
subsection)
\begin{equation}
\Psi(z_1,z_2,..,z_N; \hat y)=\frac{1}{\sqrt{(N-M)!M! L^N
N^{M}}}\det[e^{i k_i z_j}] \det[e^{i\frac{2\pi}{N} \kappa_i y_j}]
(-1)^y . \label{Psieq}
\end{equation}
Here and later we denote a sign  factor
\begin{eqnarray}
(-1)^y=\prod_{N \geq i>j \geq 1}Sign(y_i-y_j).
\end{eqnarray}
One should note, that second determinant has a size $M\times M,$
and depends only on $y_1, ... , y_M.$ Dependence of wave function
on $y_{M+1}, ... , y_{N} $ comes only through sign prefactor. For
each particular set of $y_1, ... , y_M$ there are $(N-M)!$
different configurations of  $y_{M+1}, ... , y_{N},$ for which
wave function only changes its sign depending on relative
positions of $y_{M+1}, ... , y_{N}.$

To calculate correlation function we should be able to calculate a
product of wavefunctions at the points
\begin{equation}
x_1=0, x'_1 =\xi , x'_2=x_2, .... , x'_N = x_N.
\label{xconditions}
\end{equation}
Let  $Z$  be is an ordered set for $x_i$ variables:
\begin{equation}
Z=\{z_1=0\leq z_2 \leq ... \leq z_N\leq L\}.
\label{zset}
\end{equation}
If we denote an ordered set for $x'_i$ variables  as  $Z',$ then
using (\ref{xconditions}) one  can conclude that $Z'$ is obtained
from $Z$  by removing $z_1=0,$ inserting an extra coordinate
$z'_d=\xi,$ and  shifting variables which are to the left of it:
\begin{equation}
Z'=\{0 \leq z'_1=z_2 \leq z'_2 = z_3\leq ...\leq z'_{d-1}= z_{d} \leq z'_d=\xi \leq z'_{d+1}=z_{d+1}  \leq... \leq z'_N=z_N\leq L\}.
\label{zprimeset}
\end{equation}
"Spin" states $\hat y$ and $ \hat y'$ are connected by
\begin{eqnarray}
y_1=1, y'_1=d, \nonumber \\ y'_i=y_{i}-1,\; \mbox{for}\;1< y_i
\leq d ,\nonumber \\ y'_i=y_{i} ,  \;\mbox{for} \; d < y_i.
\label{yprimeset}
\end{eqnarray}
Correlation function can be written as
\begin{eqnarray}
\rho^b(0,\xi)=M \int\Psi(0,x_2,..,x_N)\Psi^{\dagger}(\xi,
x_2...,x_N) dx_2 ... dx_N=\nonumber \\ \sum_{d=1}^{N} \sum_{\hat
y} \int \left(\frac{(-1)^y(-1)^{y'}}{(N-M)!(M-1)! L^N N^{M}}\det[e^{i k_i z_j}]
\det[e^{i \frac{2\pi}{N}\kappa_i y_j}]
 \det[e^{- i k_i z'_j}] \det[e^{-i \frac{2\pi}{N} \kappa_i y'_j}] \right) dz_2 ... dz_N,
 \label{psipsidaggerint}
\end{eqnarray}
where integration over $dz_i$ and summation over $\hat y$ are done
subject to constraints (\ref{zprimeset})-(\ref{yprimeset}). One
can observe now, that limits of integration in (\ref{zprimeset})
depend only on $\xi$ and $d.$ These limits are independent of
$\hat y,$ and function under integral factorizes into
$z-$dependent and $\hat y-$dependent parts. Similarly, summation
over $\hat y$ doesn't depend on precise values of $\xi$ or $z_i,$
but the dependence comes through $d.$ Therefore, density matrix
can be written as
\begin{equation}
\rho^b(0,\xi)=\frac{1}{(N-M)!(M-1)! L^N N^{M}}\sum_{d=1}^{N}
I(d,\xi) S^b(d), \label{sumoverd}
\end{equation}
where $I(d,\xi)$ is a an integral
\begin{equation}
I(d,\xi)=\int \det[e^{i k_i z_j}] \det[e^{-i k_i z'_j}]dz_2...dz_N
\label{Idxidef}
\end{equation}
subject to constraints (\ref{zprimeset}), and  $S^b(d)$ is an
expectation value of a translation operator over a symmetrized
Slater determinant wavefunction:
\begin{equation}
S^b(d)=\sum_{\hat y} \det[e^{i \frac{2\pi}{N}\kappa_i
y_j}]\det[e^{-i\frac{2\pi}{N} \kappa_i y'_j}](-1)^y(-1)^{y'}.
\end{equation}
Normalization can be determined using the following argument: if
$\xi=0,$ then only contribution from $d=1$ does not vanish.  One
can calculate $I(1,0)=N L^{N-1}$ and $S^b(1)=(N-M)!(M)! N^{M-1},$
since these follow from normalizations of orbital and "spin"
wavefunctions. Since we want $\rho^b(0,0)= M/L,$ we can fix the
normalization prefactor in (\ref{Psieq}).

\subsubsection{Calculation of a many-body integral $I(d,\xi)$}
Let's describe the calculation of an integral $I(d,\xi)$. From now
on we will assume that $L=1.$ First, since $k_i$ are equidistant
wave vectors (\ref{kset}), one can use Vandermonde formula to
simplify the determinants:
\begin{eqnarray}
\det[e^{i k_l z_j}]=e^{-i \pi(N-1)(z_1+...+z_N)} \det[e^{i 2\pi
(l-1) z_j}]= e^{-i \pi(N-1)(z_1+...+z_N)} \prod_{j1<j2}(e^{i 2\pi
z_{j2}}-e^{i 2 \pi  z_{j1}}) , l = \{1,...,N\},\nonumber \\
\det[e^{-i k_l z'_j}]=e^{i \pi(N-1)(z'_1+...+z'_N)} \det[e^{-i2\pi
(l-1) z'_j}]= e^{i \pi(N-1)(z'_1+...+z'_N)}
\prod_{j1<j2}(e^{-i2\pi  z'_{j2}}-e^{-i2\pi  z'_{j1}}),l =
\{1,...,N\}. \label{vander1}
\end{eqnarray}
Using this representation, the fact that $z_1=0$ and (\ref{zprimeset}), one can rewrite  these $N \times N$ determinants as a product of $(N-1)\times (N-1)$
determinant and a prefactor:
\begin{eqnarray}
\det[e^{i k_l z_j}]=e^{-i \pi(N-1)(t_1+...+t_{N-1})} \det[e^{i
2\pi (l-1) t_l}] \prod_{i=1}^{N-1}(e^{i 2\pi  t_{i}}-1), l =
\{1,...,N-1\}\nonumber \\ \det[e^{i k_l z'_j}]=(-1)^{d-1} e^{i
\pi(N-1)(\xi+ t_1+...+t_{N-1})} \det[e^{-i 2\pi (l-1) t_l}]
\prod_{i=1}^{N-1}(e^{-i 2\pi  t_{i}}-e^{-i 2\pi  \xi}),l =
\{1,...,N-1\}, \label{vander2}
\end{eqnarray}
where we introduced $N-1$ variables of integration $t_i,$ so that
\begin{eqnarray}
t_i=z_{i+1}.
\end{eqnarray}
Factor $(-1)^{d-1}$ arises since $z'_d=\xi,$ and to write
(\ref{vander2}) we changed signs of $d-1$ terms in
(\ref{vander1}). Integration subspace is defined as
\begin{equation}
 \{0 \leq t_1\leq... \leq t_{d-1} \leq \xi \leq t_{d} \leq... \leq t_{N-1}\leq 1\}
\end{equation}
One can extend this subspace  as follows:
\begin{equation}
 T=\{0 \leq t_1, ... , t_{d-1} \leq \xi \leq t_{d}, ..., t_{N-1}\leq1 \}.
 \label{t}
\end{equation}
Indeed, expression under integral doesn't change, when $t_i<\xi$ and $t_j<\xi$  change their positions(similarly for $t_i>\xi$ and $t_j>\xi$ ), so this
extension just adds prefactor $1/((d-1)!(N-d)!).$ Finally, we have
\begin{equation}
I(d,\xi)=\frac{( -1)^{d-1} e^{i
\pi(N-1)\xi}}{(d-1)!(N-d)!}\int_{t_i\subset T}\det[e^{i 2\pi (l-1)
t_l}]\det[e^{-i 2\pi (l-1) t_l}]
  \prod_{i=1}^{N-1}(e^{i 2\pi  t_{i}}-1)(e^{-i 2\pi  t_{i}}-e^{-i 2\pi  \xi}) dt_1...dt_{N-1}.
\end{equation}
At this point we use a trick from \cite{Lenard}, where Toeplitz determinant representation for strongly interacting bose gas was derived.
Lets expand determinants under integrals using permutation formula for determinants:
\begin{eqnarray}
I(d,\xi) =\frac{(-1)^{d-1} e^{i \pi(N-1)\xi}}{(d-1)!(N-d)!}
\int_{t_i\subset T} \sum_{P\subset S_{N-1}} \sum_{P'\subset
S_{N-1}} (-1)^P (-1)^{P'} \times \\
\prod_{i=1}^{N-1} e^{i2\pi
((P_i-1)-(P'_i-1)) t_i}(e^{i 2\pi  t_{i}}-1)(e^{-i 2\pi
t_{i}}-e^{-i 2\pi  \xi}) dt_1...dt_{N-1}
\end{eqnarray}
From summation over $P, P'$ we can go to summation over $P, Q,$ where $P'= Q P.$ Also, one can remove constraints (\ref{t}) by introducing two functions
\begin{eqnarray}
f^1(\xi, t)=(e^{i 2\pi  t}-1)(e^{-i 2\pi  t}-e^{-i 2\pi  \xi})\,\,
\mbox{for}  \, t<\xi,\; 0  \,\, \mbox{otherwise}, \nonumber \\
f^2(\xi, t)=(e^{i 2\pi  t}-1)(e^{-i 2\pi  t}-e^{-i 2\pi  \xi})\,\,
\mbox{for}  \, t>\xi, \; 0 \,\, \mbox {otherwise}.
\end{eqnarray}
$I(d,\xi)$  becomes
\begin{eqnarray}
 \frac{(-1)^{d-1}e^{i \pi(N-1)\xi}}{(d-1)!(N-d)!} \sum_{P\subset S_{N-1}} \sum_{Q\subset S_{N-1}} (-1)^Q  (\prod_{i=1}^{d-1}
 \int_{0}^1 e^{i2\pi (P_i-Q_{P_i} ) t_i} f^1(\xi,t_i) dt_i)(\prod_{i=d}^{N-1} \int_{0}^1 e^{i2\pi (P_i-Q_{P_i} ) t_i} f^2(\xi,t_i) dt_i)
 \label{ggg}
\end{eqnarray}
If $f^1(\xi,t)$ and $f^2(\xi,t)$ were the same, as in
\cite{Lenard}, expression being summed wouldn't depend on $P,$ and
summation over $Q$ would give a determinant, with the same
elements along diagonals(Toeplitz determinant). In our case, for
each given $P$ the expression is $P-$ dependent, and the result
doesn't have the  Toeplitz form. However, introducing additional
"phase" variable, one can recast the expression as an integral of
some Toeplitz determinant. Desired expression has the form:
\begin{eqnarray}
I(d,\xi) =(-1)^{d-1}e^{i \pi(N-1)\xi}\int_0^{2\pi} \frac{d\varphi}{2\pi} e^{- i (d-1)\varphi}(\sum_Q (-1)^Q \prod_{i=1}^{N-1} \int_{0}^1 e^{i2\pi (i-Q_i ) c_i} (e^{i \varphi} f^1(\xi,c_i) +f^2(\xi,c_i)) dc_i),
\label{Idxi}
\end{eqnarray}
where $c_i$ is a dummy variable of integration. Integration over $\varphi$ is analogous to projection of BCS to a state with a fixed number of particles.
After integration over $\varphi$ nonzero terms appear, if in the expansion of the  product of brackets  for some $d-1$ brackets $f^1$ is chosen instead of $f^2.$ If this choice is made at brackets with
numbers $P_1, ..., P_{d-1}$ then contribution from such a choice exactly corresponds to a term in (\ref{ggg}). However, each choice of brackets corresponds
to $(N-d)!(d-1)!$ different permutations, and this cancels the same  combinatoric factor in the denominator of (\ref{ggg}).

Summation over $Q$ is nothing but a determinant, and finally we have
\begin{equation}
I(d,\xi) =(-1)^{d-1}e^{i \pi(N-1)\xi}\int_0^{2\pi}
\frac{d\varphi}{2\pi} e^{- i (d-1)\varphi}
\det\left[\begin{array}{cccc} c_0 (\varphi)& c_1(\varphi)&  ...
&c_{N-2}(\varphi)\\
  c_{-1}(\varphi) & c_0(\varphi)& ... &c_{N-3}(\varphi)\\
  ... & ... & ... & ...\\
  c_{-(N-2)}(\varphi) & c_{-(N-3)}(\varphi)& ... &c_0(\varphi)
  \end{array}\right],
  \label{Idxifinal}
\end{equation}
where
\begin{equation}
c_j(\varphi)=\int_0^1  e^{i j x}(e^{i\varphi}f^1(\xi, x) +f^2(\xi, x))dx
\end{equation}
Expression in (\ref{Idxifinal}) without an integral over $\varphi$
is a generating function of $I(d,\xi)$ with the weights
$e^{i(\varphi-\pi)(d-1)}$, and integration over $\varphi$ extracts
a particular term out of this generating function.

What we achieved in this section is to represent a complicated
$N-1$ fold integral as an integral over one phase variable, which
can be done numerically
 in a polynomial time over $N.$

\subsubsection{Calculation of $S^b(d)$}
Calculation  of $S^b(d)$ is very similar in spirit to calculation
of the previous subsection. Integration over $x_i$ corresponds to
summation over $y_i,$ and $\xi$ corresponds to $d.$ Final result
is  a determinant of some matrix. Due to the shift operator
(\ref{yprimeset}) this determinant does not  have a Toeplitz form,
but it is not important for a numerical evaluation.

We need to calculate
\begin{equation}
S^b(d)=\sum_{\hat y} \det[e^{i \frac{2\pi}{N}\kappa_i
y_j}]\det[e^{-i\frac{2\pi}{N} \kappa_i y'_j}](-1)^y(-1)^{y'},
\label{sdprelim}
\end{equation}
where $\kappa_i$ is a set (\ref{kappaset}).
Definition of $\hat y'$ according to (\ref{yprimeset}) can be rewritten as
\begin{eqnarray}
y_1=1 , y'_1=d, \nonumber \\
y'_i = y_i +\frac{Sign(y_i - d) - 1}2, i =\{2, ... , N\},
\end{eqnarray}
where
\begin{eqnarray}
Sign(x)=1, \, x>0 ,\nonumber \\ Sign(x)=-1, \, x\leq 0.
\end{eqnarray}
 Sign prefactor in (\ref{sdprelim}) can be rewritten as
 \begin{eqnarray}
 (-1)^y(-1)^{y'}=\prod_{i>j}Sign(y_i-y_j)\prod_{i>j}Sign(y'_i-y'_j)=\prod_{i=2}^{N}Sign(y_i -d)=(-1)^{d-1}.
 \label{signprefactor}
 \end{eqnarray}
 We see, that (\ref{sdprelim}) depends only on $y_1, ... , y_M,$  so from now on we will consider a summation in $y_1 , ... , y_M$ variables.
 Summation over $y_{M+1}, ..., y_N$ gives a trivial combinatorial prefactor $(N-M)!.$ Furthermore, we can extend possible values of $y_1 , ... , y_M$ to
 $y_i=y_j , i\neq j,$ since for such configurations first determinant in (\ref{sdprelim}) is $0,$ and they don't change the value of $S^b(d):$
\begin{equation}
\hat y =\{ 1 \leq y_2, y_3, ... , y_M\leq N\}.
\label{hatyset}
\end{equation}

 Lets use the fact that $\kappa_i$ is a set of equidistant numbers (\ref{kappaset}), and rewrite determinants using Vandermonde formula, similar to (\ref{vander1}):
 \begin{eqnarray}
 \det[e^{i \frac{2\pi}{N}\kappa_i y_j}]=  e^{i \frac{2\pi}{N}(-(M-1)/2+N/2)(1+...+y_M)} \det[e^{i \frac{2\pi}{N} (l-1) y_j}]=\nonumber \\
 e^{i \frac{2\pi}{N}(-(M-1)/2+N/2)(1+...+y_M)}\prod_{j1<j2}(e^{i \frac{2\pi}{N}  y_{j2}}-e^{i  \frac{2\pi}{N}  y_{j1}}) , l = \{1,...,M\}\nonumber,\\
 \det[e^{-i \frac{2\pi}{N}\kappa_i y'_j}]= e^{-i \frac{2\pi}{N}(-(M-1)/2+N/2)(d+...+y'_M)} \det[e^{i \frac{2\pi}{N} (l-1) y'_j}]=\nonumber \\
 e^{-i \frac{2\pi}{N}(-(M-1)/2+N/2)(d+...+y'_M)}\prod_{j1<j2}(e^{i \frac{2\pi}{N}  y'_{j2}}-e^{i  \frac{2\pi}{N}  y'_{j1}}) , l = \{1,...,M\}.
 \label{vander3}
 \end{eqnarray}
 For simplicity of notations later, lets introduce $t_i=y_{i+1}, t'_i=y'_{i+1}, i=\{1, ... , M-1\}.$
  Analogously to (\ref{vander2}), we extract a  determinant of $(M-1)\times(M-1)$ matrix  out of Vandermonde
 product:
 \begin{eqnarray}
  \det[e^{i \frac{2\pi}{N}\kappa_i y_j}]=e^{i \frac{2\pi}{N}(-(M-1)/2+N/2)(1+t_1...+t_{M-1})} \det[e^{i \frac{2\pi}{N} (l-1) t_j}]
  \prod _{i=1}^{M-1}(e^{i \frac{2\pi}{N}  t_{i}}-e^{i  \frac{2\pi}{N} 1 }) , l = \{1,...,M-1\} ,\nonumber\\
  \det[e^{-i \frac{2\pi}{N}\kappa_i y'_j}]=e^{-i \frac{2\pi}{N}(-(M-1)/2+N/2)(d+t'_1...+t'_{M-1})} \det[e^{-i \frac{2\pi}{N} (l-1) t'_j}]
  \prod _{i=1}^{M-1}(e^{-i \frac{2\pi}{N}  t_{i}}-e^{-i  \frac{2\pi}{N} d }) , l = \{1,...,M-1\} .
  \label{vander4}
 \end{eqnarray}
 At this point we need to represent the subspace of summation (\ref{hatyset}) as a sum over $M$ inequivalent partitions, similar to representation (\ref{sumoverd}):
 \begin{equation}
 S^b(d)=\sum_{r=1}^{M} \frac{(N-M)!(M-1)!}{(r-1)!(M-r)!}S^b(d, r),
 \label{sdrepr}
 \end{equation}
 where $S^b(d, r)$ is a result of summation  in the $T_r$ subspace:
 \begin{equation}
 T_r =\{1\leq t_1, ... , t_{r-1} \leq d < t_r, ... , t_{M-1}\leq N\}.
 \end{equation}
Note, that $S^b(d,r) =0 $  for $r>d,$ since in this case two of
$t_1, ..., t_{r-1}$ should coincide, and wavefunction becomes $0$.
Calculation of $S^b(d,r)$ is very similar to calculation of
$I(\xi,d).$  Let's expand the determinants (\ref{vander4}) using
permutations:
\begin{eqnarray}
S^b(d,r)=(-1)^{d-1}e^{i \frac{2 \pi}{N}(-(M-1)/2 + N/2) (r-d)}
\sum_{P\subset S_{M-1}} \sum_{P'\subset S_{M-1}}(-1)^P (-1)^{P'}
\times \nonumber \\
 \prod_{i=1}^{r-1} ( \sum_{t_i=1}^{d} e^{i\frac{2\pi}{N} ((P'_i-1)t_i-(P_i-1) (t_i-1))}(e^{i \frac{2\pi}{N}t_i}-e^{i \frac{2\pi}{N}})(e^{-i \frac{2\pi}{N}  (t_{i}-1)}-e^{-i \frac{2\pi}{N}  d}))
 \times \nonumber \\
 \prod_{i=r}^{M-1}(\sum_{t_i=d+1}^{N}  e^{i\frac{2\pi}{N} ((P'_i-1)-(P_i-1)) t_i}(e^{i \frac{2\pi}{N}t_i}-e^{i \frac{2\pi}{N}})(e^{-i \frac{2\pi}{N}  t_{i}}-e^{-i \frac{2\pi}{N} d}))
\end{eqnarray}
 From summation over $P, P'$ we can go to summation over $P, Q,$ where $P'= Q P.$ Also, one can analytically perform summation over $t_i$ in each
 of the brackets, since it is a combination of geometrical progressions(this is analogous to  integration over $t_i$ variables in previous subsection):
  \begin{eqnarray}
 S^b(d,r)=(-1)^{d-1}e^{i \frac{2 \pi}{N}(-(M-1)/2 + N/2) (r-d)} \sum_{P\subset S_{M-1}} \sum_{Q\subset S_{M-1}}(-1)^Q
 \prod_{i=1}^{r-1}c^1(d,Q_{P_i}, P_i)
 \prod_{i=r}^{M-1}c^2(d,Q_{P_i}, P_i),
 \label{sdr}
 \end{eqnarray}
 where
 \begin{eqnarray}
 c^1(d, j, l)=e^{i\frac{2\pi}{N}(l-1)}\sum_{t=1}^{t=d}e^{i\frac{2\pi}{N}(j-l)  t}(e^{i \frac{2\pi}{N}d}-e^{i \frac{2\pi}{N}})(e^{-i \frac{2\pi}{N}  (t-1)}-e^{-i \frac{2\pi}{N}  d})\nonumber, \\
 c^2(d, j, l) = \sum_{t=d+1}^{t=N}e^{i\frac{2\pi}{N}(j-l)  t}(e^{i \frac{2\pi}{N}d}-e^{i \frac{2\pi}{N}})(e^{-i \frac{2\pi}{N} t}-e^{-i \frac{2\pi}{N}  d})
 \end{eqnarray}
are independent of $r.$
 At this point, we can use the "phase" variable integration trick to get rid of summation over $P,$ and then represent summation over $Q$ as a determinant:
 \begin{eqnarray}
 S^b(d,r)=(r-1)!(M-r)!(-1)^{d-1}e^{i \frac{2 \pi}{N}(-(M-1)/2 + N/2) (r-d)}\times  \nonumber \\
 \int_0^{2\pi} \frac{d\psi}{2\pi} e^{- i (r-1)\psi}
 \det\left[\begin{array}{cccc}
c(\psi,1,1)& c(\psi,2,1)&  ... &c(\psi,M-1,1)\\
  c(\psi,1,2) & c(\psi,2,2)& ... &c(\psi,M-1,2)\\
  ... & ... & ... & ...\\
  c(\psi,1,M-1) & c(\psi,2,M-1)& ... &c(\psi,M-1,M-1)
  \end{array}\right],
  \label{sdr2}
 \end{eqnarray}
  where
  \begin{equation}
  c(\psi,j,l)=e^{i\psi}c^1(d, j,l)+c^2(d, j, l).
  \end{equation}
  We can analytically perform summation over $r$  in (\ref{sdrepr}), since  the determinant and $c(\psi,j,l)$  are independent of $r:$
  \begin{eqnarray}
  S^b(d)=\sum_{r=1}^{M} \frac{(N-M)!(M-1)!}{(r-1)!(M-r)!}S^b(d, r)=(N-M)!(M-1)!(-1)^{d-1}e^{-i \frac{2 \pi}{N}(-(M-1)/2 + N/2) (d-1)}\times \nonumber \\
\int_0^{2\pi} \frac{d\psi}{2\pi}
    (\sum_{r=1}^{M}e^{i (\frac{2 \pi}{N}(-(M-1)/2 + N/2) -\psi)(r-1)})
\det\left[\begin{array}{cccc} c(\psi,1,1)& c(\psi,2,1)&  ...
&c(\psi,M-1,1)\\
  c(\psi,1,2) & c(\psi,2,2)& ... &c(\psi,M-1,2)\\
  ... & ... & ... & ...\\
  c(\psi,1,M-1) & c(\psi,2,M-1)& ... &c(\psi,M-1,M-1)
  \end{array}\right].
  \label{sdfinal}
  \end{eqnarray}

  Expansion of the determinant (\ref{sdfinal}) in a series over $e^{i \psi}$ has terms up to $e^{i(M-1) \psi}:$
  \begin{equation}
  \det(\psi)=\sum_{n=0}^{M-1}f_n e^{i n \psi}.
  \end{equation}
 Summation over $r$  and integration over $\psi$ lead to
 \begin{eqnarray}
 \int_0^{2\pi} \frac{d\psi}{2\pi}(\sum_{r=1}^{M}e^{i (\frac{2 \pi}{N}(-(M-1)/2 + N/2) -\psi)(r-1)})\det(\psi)=
 \int_0^{2\pi} \frac{d\psi}{2\pi}\sum_{n=0}^{M-1}\sum_{r=1}^{M}f_n e^{i (\frac{2 \pi}{N}(-(M-1)/2 + N/2) -\psi)(r-1)+i\psi n}=\nonumber\\
 \sum_{n=0}^{M-1}f_n e^{i \frac{2 \pi}{N}(-(M-1)/2 + N/2)n}=\det(\frac{2 \pi}{N}(-(M-1)/2 + N/2)).
 \end{eqnarray}
Finally, if we introduce a notation $\psi_0=2 \pi(-(M-1)/2 + N/2)/N,$
  \begin{eqnarray}
  S^b(d)=(N-M)!(M-1)!(-1)^{d-1}e^{-i \frac{2 \pi}{N}(-(M-1)/2 + N/2) (d-1)}\times \nonumber \\
  \det\left[\begin{array}{cccc} c(\psi_0,1,1)& c(\psi_0,2,1)&  ...
&c(\psi_0,M-1,1)\\
  c(\psi_0,1,2) & c(\psi_0,2,2)& ... &c(\psi_0,M-1,2)\\
  ... & ... & ... & ...\\
  c(\psi_0,1,M-1) & c(\psi_0,2,M-1)& ... &c(\psi_0,M-1,M-1)
  \end{array}\right].
  \label{sdfinalfinal}
  \end{eqnarray}

\subsection{Fermi-Fermi correlation function}
\label{ff}
 Calculation of fermionic correlation function closely reminds the calculation of Bose-Bose correlation function, so we will be
 sufficiently sketchy in our derivation.  First, one splits integration into integration over orbital coordinates $z_i$ from the set
\begin{equation}
Z=\{0\leq z_1 \leq z_2 \leq ...  \leq z_N \leq L\},
\label{fermizhatset}
\end{equation}
 and summation  over "spin" variables.
 Integration over orbital variables is absolutely identical to the Bose-Bose case, the difference comes only from "spin" part $S^f(d):$
 \begin{eqnarray}
 \rho^f(0,\xi)=(N-M) \int\Psi(x_1,x_2, ... ,0)\Psi^{\dagger}(x_1, x_2...,\xi) dx_1 ... dx_{N-1}=\nonumber \\
\sum_{d=1}^{N} \sum_{\hat y} \int \left(\frac{(-1)^y(-1)^{y'}}{(N-M-1)!M! L^N
N^{M}}\det[e^{i k_i z_j}] \det[e^{i \frac{2\pi}{N}\kappa_i y_j}]
 \det[e^{- i k_i z'_j}] \det[e^{-i \frac{2\pi}{N} \kappa_i y'_j}] \right) dz_2 ... dz_N=\nonumber\\
 \frac{1}{(N-M-1)!M! L^N N^{M}}\sum_{d=1}^{N} I(d,\xi) S^f(d),
 \label{fermipsipsidaggerint}
 \end{eqnarray}
where $I(d,\xi)$ is given by (\ref{Idxifinal}), and
\begin{eqnarray}
S^f(d)=\sum_{\hat y} \det[e^{i \frac{2\pi}{N}\kappa_i
y_j}]\det[e^{-i\frac{2\pi}{N} \kappa_i y'_j}](-1)^y(-1)^{y'}.
\label{sfd}
\end{eqnarray}
 In (\ref{sfd}) $\hat y'$ and $\hat y$ are related by
 \begin{eqnarray}
y_N=1 , y'_N=d, \nonumber \\ y'_i = y_i +\frac{Sign(y_i - d) -
1}2, i =\{1, ... , N-1\} \label{fermiyprimeset}.
 \end{eqnarray}
 Similarly to (\ref{signprefactor}) sign prefactor can be rewritten as
 \begin{eqnarray}
 (-1)^y(-1)^{y'}=\prod_{i>j}Sign(y_i-y_j)\prod_{i>j}Sign(y'_i-y'_j)=(-1)^{N-1}\prod_{j=1}^{N-1}Sign(d-y_j)=(-1)^{d-1}.
 \end{eqnarray}

 We see , that (\ref{sfd}) depends only on $y_1, ... , y_M,$  so from now on we will consider a summation in $y_1 , ... , y_M$ variables.
 Summation over $y_{M+1}, ..., y_{N-1}$ gives a trivial combinatorial prefactor $(N-M-1)!.$ Furthermore, we can extend possible values of $y_1 , ... , y_M$ to
 $y_i=y_j , i\neq j,$ since for such configurations first determinant in (\ref{sfd}) is $0,$ and they don't change the value of $S^f(d):$
\begin{equation}
\hat y =\{ 2 \leq y_1, y_2, ... , y_M\leq N\}.
\label{fermihatyset}
\end{equation}

 We can to represent the subspace of summation (\ref{fermihatyset}) as a sum  of $M+1$ inequivalent partitions, similar to representation (\ref{sdrepr}):
 \begin{equation}
 S^f(d)=\sum_{r=1}^{M+1} \frac{(N-M-1)!(M)!}{(r-1)!(M-r+1)!}S^f(d, r),
 \label{sfdrrepr}
 \end{equation}
 where $S^f(d, r)$ is a result of summation  in the $T_r$ subspace:
 \begin{equation}
 T_r =\{2\leq t_1, ... , t_{r-1} \leq d < t_r, ... , t_{M}\leq N\}.
 \end{equation}
 Product of two determinants in (\ref{sfd}) is rewritten as
  \begin{eqnarray}
 \det[e^{i \frac{2\pi}{N}\kappa_i y_j}]\det[e^{-i \frac{2\pi}{N}\kappa_i y'_j}]=  e^{i \frac{2\pi}{N}(-(M-1)/2+N/2)(r-1)} \det[e^{i \frac{2\pi}{N} (l-1) y_j}]
 \det[e^{-i \frac{2\pi}{N} (l-1) y'_j}], l = \{1,...,M\}.
 \label{fermivander}
 \end{eqnarray}
 We can expand the determinants (\ref{fermivander}) using permutations:
\begin{eqnarray}
S^f(d,r)=(-1)^{d-1}e^{i \frac{2 \pi}{N}(-(M-1)/2 + N/2) (r-1)} \sum_{P\subset S_{M}} \sum_{P'\subset S_{M}}(-1)^P (-1)^{P'}
\times \nonumber \\
 \prod_{i=1}^{r-1} ( \sum_{t_i=2}^{d} e^{i\frac{2\pi}{N} ((P'_i-1)t_i-(P_i-1) (t_i-1))})
  \prod_{i=r}^{M}(\sum_{t_i=d+1}^{N}  e^{i\frac{2\pi}{N} ((P'_i-1)-(P_i-1)) t_i}).
\end{eqnarray}
 From summation over $P, P'$ we can go to summation over $P, Q,$ where $P'= Q P.$ Also , one can analytically perform summation over $t_i$ in each
 of the brackets, since it is a  geometrical progression.
  \begin{eqnarray}
 S^f(d,r)=(-1)^{d-1}e^{i \frac{2 \pi}{N}(-(M-1)/2 + N/2) (r-1)} \sum_{P\subset S_{M}} \sum_{Q\subset S_{M}}(-1)^Q
 \prod_{i=1}^{r-1}c_f^1(d,Q_{P_i}, P_i)
 \prod_{i=r}^{M}c_f^2(d,Q_{P_i}, P_i),
 \label{fermisdr}
 \end{eqnarray}
 where
 \begin{eqnarray}
 c^1_f(d, j, l) = e^{i\frac{2\pi}{N}(l-1)}\sum_{t=2}^{t=d}e^{i\frac{2\pi}{N} (j-l)  t}\nonumber, \\
 c^2_f(d, j, l) = \sum_{t=d+1}^{t=N}e^{i\frac{2\pi}{N}(j-l)  t}
 \end{eqnarray}
are independent of $r.$
 At this point, we can use the "phase" variable integration trick to get rid of summation over $P,$ and then represent summation over $Q$ as a determinant:
 \begin{eqnarray}
 S^f(d,r)=(r-1)!(M-r+1)!(-1)^{d-1}e^{i \frac{2 \pi}{N}(-(M-1)/2 + N/2) (r-1)}\times  \nonumber \\
 \int_0^{2\pi} \frac{d\psi}{2\pi} e^{- i (r-1)\psi}
 \det\left[\begin{array}{cccc}
c_f(\psi,1,1)& c_f(\psi,2,1)&  ... &c_f(\psi,M,1)\\
  c_f(\psi,1,2) & c_f(\psi,2,2)& ... &c_f(\psi,M,2)\\
  ... & ... & ... & ...\\
  c_f(\psi,1,M) & c_f(\psi,2,M)& ... &c_f(\psi,M,M)
  \end{array}\right],
  \label{fermisdr2}
 \end{eqnarray}
  where
  \begin{equation}
  c_f(\psi,j,l)=e^{i\psi}c^1_f(d, j,l)+c^2_f(d, j, l).
  \end{equation}
  We can analytically perform summation over $r$ in (\ref{sfdrrepr}), since the  form of the determinant and $c_f(\psi,j,l)$  are independent of $r,$ and $r-$dependent combinatorial prefactor
  cancels:
  \begin{eqnarray}
  S^f(d)=\sum_{r=1}^{M+1} \frac{(N-M-1)!M!}{(r-1)!(M-r+1)!}S^f(d, r)=(N-M-1)!(M)!(-1)^{d-1}\times\nonumber\\
 \int_0^{2\pi} \frac{d\psi}{2\pi}
    \frac{e^{i (\frac{2 \pi}{N}(-(M-1)/2 + N/2) -\psi)(M+1)}-1}{e^{i (\frac{2 \pi}{N}(-(M-1)/2 + N/2) -\psi)}-1}\det\left[\begin{array}{cccc}
c_f(\psi,1,1)& c_f(\psi,2,1)&  ... &c_f(\psi,M,1)\\
  c_f(\psi,1,2) & c_f(\psi,2,2)& ... &c_f(\psi,M,2)\\
  ... & ... & ... & ...\\
  c_f(\psi,1,M) & c_f(\psi,2,M)& ... &c_f(\psi,M,M)
  \end{array}\right].
  \label{fermisdfinal}
  \end{eqnarray}
  Analogously to the case of bosons, integration over $\psi$ is equivalent to substitution $\psi_0=2 \pi(-(M-1)/2 + N/2)/N$ to the determinant, and
  the final expression is
  \begin{eqnarray}
  S^f(d)=(N-M-1)!(M)!(-1)^{d-1}
  \det\left[\begin{array}{cccc}
  c_f(\psi_0,1,1)& c_f(\psi_0,2,1)&  ...&c_f(\psi_0,M,1)\\
  c_f(\psi_0,1,2) & c_f(\psi_0,2,2)& ... &c_f(\psi_0,M,2)\\
  ... & ... & ... & ...\\
  c_f(\psi_0,1,M) & c_f(\psi_0,2,M)& ... &c_f(\psi_0,M,M)
  \end{array}\right].
  \label{fermisdfinalfinal}
  \end{eqnarray}

\subsection{Numerical evaluation of correlation functions and Luttinger parameters}

Using results of the previous sections, one can evaluate
correlation functions on a ring numerically and extract both
long-range and short range behavior of correlation functions.
Calculation of all determinants requires polynomial time in their size,
and systems of up to $N=100$ atoms can be easily investigated on a desktop PC.
Fourier
transform of correlation function is  an occupation number $n(k),$
which can be measured directly in time-of-flight
experiments\cite{Paredes} or using Bragg spectroscopy\cite{Bragg}.
Recently, long distance correlation functions of the model under consideration
have been investigated based on conformal field theory (CFT)
arguments\cite{Frahm}. Our determinant representations for strongly
interacting mixture can be used to obtain these correlation functions at
all distances, and compare their large distance asymptotic behavior
with predictions of CFT.

\begin{figure}
\psfig{file=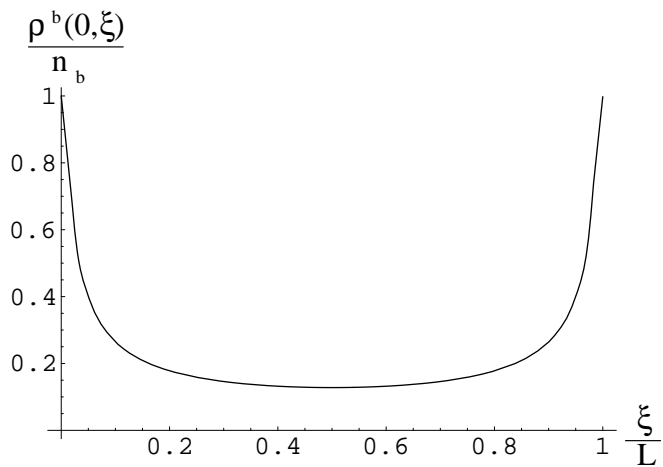} \caption{ \label{corrfunc} Normalized
bose-bose correlation function on a circle as a function of  the distance $\xi$ (here $n_b$ is bose density). Due to periodic boundary conditions correlation function is periodic with a period $L,$ where $L$ is the size of the system. Numerical evaluation is done for  $M=15,N=30.$}
\end{figure}

 In fig \ref{corrfunc} we show numerically evaluated Bose-Bose correlation function for $M=15, N=30.$ Since we used periodic boundary conditions,
 correlation function is periodic in $\xi.$
To extract universal long-distance correlation functions from our calculation, one has to fit the numerical results using general Luttinger liquid asymptotic behavior. In the thermodynamic limit long range behavior is
\begin{equation}
\rho^b(0,\xi)\sim |\xi|^{-1/(2K_b)},
\end{equation}
 where $K_b$ is a bosonic Luttinger Liquid parameter.  This formula is valid, if $\xi$ is bigger then any  non-universal short-range scale of the model.
 In our case, such short-range scale is given by the interbosonic distance, which is $L/M.$
 For a finite size system, general arguments of conformal invariance\cite{Tsvelik, Cazalilla} imply that correlation function has the form
 \begin{equation}
 \rho^b(0,\xi)\sim \frac{1}{|\sin{\frac{\pi \xi}{L}}|^{1/(2K_b)}}.
 \label{fitfunc}
 \end{equation}

We fitted numerically obtained correlation functions with (\ref{fitfunc}),
and results coincide with the formula
\begin{equation}
K_b=\frac1{(\alpha-1)^2-1},
\end{equation}
 obtained in \cite{Frahm} based on CFT arguments.
One can see subleading oscillations in the numerical evaluation, but their
quantitative analysis would require more numerical effort. Fourier transform of
$\rho^b(0,\xi)$ is a monotonously decreasing function, which has a singularity at $k=0,$
governed by Luttinger liquid parameter $K_b:$
\begin{equation}
n^b(k)\sim |k|^{-1+1/(2K_b)}\; \mbox{for}\; k\rightarrow 0.
\label{nbksingularity}
\end{equation}

\begin{figure}
\psfig{file=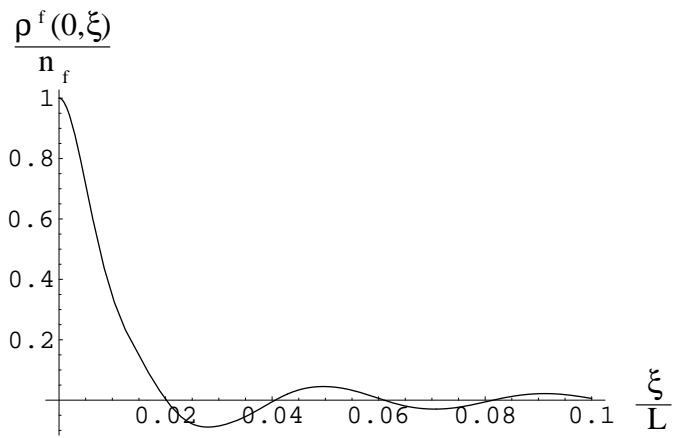} \caption{ \label{fermicorrfig} Normalized
fermi-fermi correlation function on a circle as a function of  the distance $\xi$ (here $n_f$ is fermi density and $L$ is the size of the system). Oscillations are reminiscent of Friedel
oscillations of the ideal fermi gas, and their large distance decay is controlled by Luttinger liquid behavior. Numerical evaluation is done for  $M=51,N=100.$}
\end{figure}

 Fermionic correlation functions can also be obtained using the results of the previous
section, and space dependence of a typical correlation function is presented in figure
\ref{fermicorrfig}. Oscillations are reminiscent of Friedel
oscillations of the ideal fermi gas. Their large distance decay is controlled by Luttinger liquid behavior.

\begin{figure}
\psfig{file=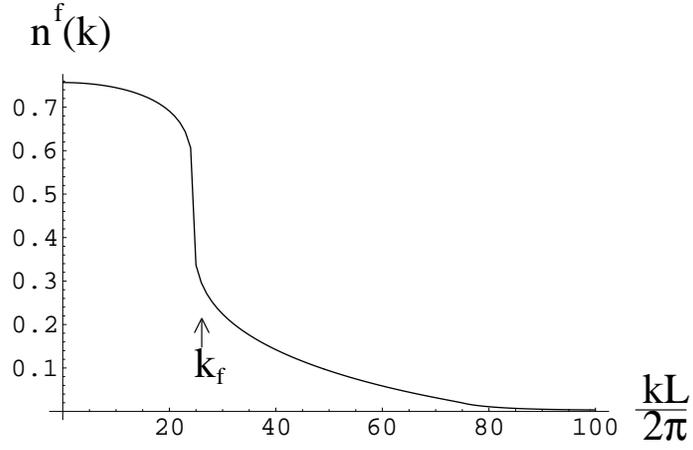} \caption{ \label{fermink} Fourier
transform of the Fermi-Fermi correlation function for
$M=51, N=100.$ Fermi step at $k_f$ gets smeared out by interactions,
but relative change of occupation number as $k_f$ is crossed is significant.}
\end{figure}

\begin{figure}
\psfig{file=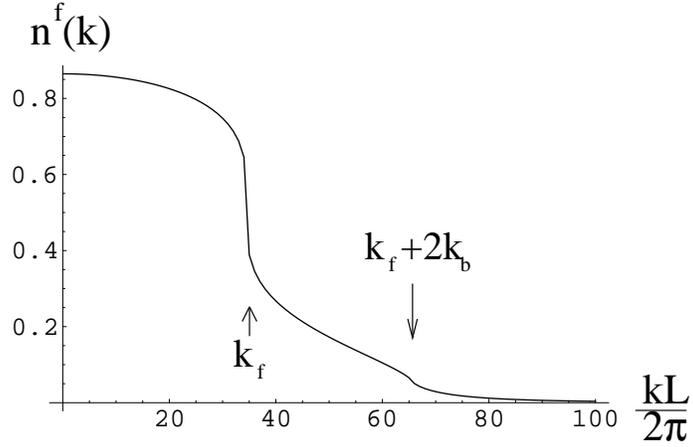} \caption{ \label{fermink1} Fourier
transform of the Fermi-Fermi correlation function for
$M=31, N=100.$ Fermi step at $k_f$ gets smeared out by interactions, and additional
discontinuity appears at $k_f+ 2 k_b.$}
\end{figure}

\begin{figure}
\psfig{file=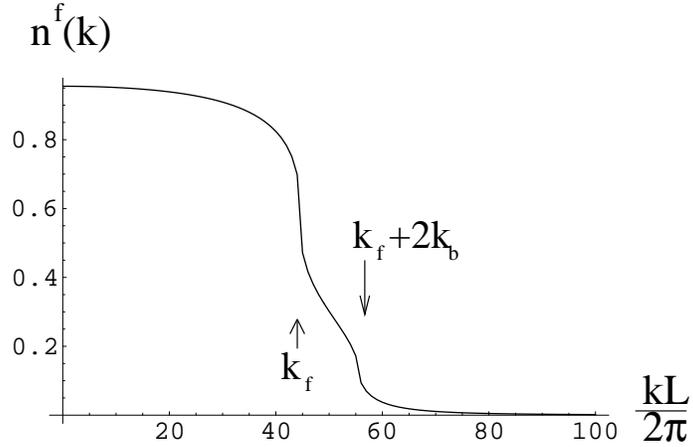} \caption{ \label{fermink2} Fourier
transform of the Fermi-Fermi correlation function for
$M=11, N=100.$ Discontinuity at $k_f+ 2 k_b$ gets stronger as $M/N$ decreases.}
\end{figure}

One can investigate Fourier transform of the correlation function, which is an occupation number, and results for different boson fractions are shown in figs. \ref{fermink}-\ref{fermink2}. In figure \ref{fermink} densities of bosons and fermions are almost equal. Fermi step at $k_f$ gets smeared out by interactions, but relative change of occupation number as $k_f$ is crossed is significant. As boson fraction is decreasing,
the discontinuity appears at $k_f+ 2 k_b,$ and it gets stronger as $M/N$ decreases (see figs. \ref{fermink1},\ref{fermink2}). The presence of this discontinuity has been predicted in \cite{Frahm}, based on CFT arguments, and here we quantify the strength of the effect.
One should note, that discontinuity at $k_f + 2 k_b$ is a direct signature of the interactions
and its detection can serve as an unambiguous verification of our theory.

\section{Low temperature behavior in Tonks-Girardeau regime}

\label{nztcorrelations}
In the previous sections we considered density profiles and developed an
algorithm to calculate the correlation functions of the ground
state of the bose-fermi hamiltonian (\ref{initialhamiltoniansc})
in the strongly interacting regime. An important question, which
is very relevant experimentally, is the effect of finite
temperatures. In principle, one can use techniques of the thermodynamic Bethe ansatz \cite{takahashi}  to obtain free energy at nonzero temperatures as the function
of interaction strength and densities. Combined with local density approximation,
it can be used to calculate density profiles for any interaction strength.
In this section we will limit our discussion to effects of small nonzero temperatures
$T\ll E_f=(\pi \hbar n)^2/(2m)$  only for strongly interacting regime.  We will show the evolution of the density profile (see fig. \ref{nztdensity}) in a harmonic trap and calculate
the correlation functions under periodic boundary conditions.
The effect of nonzero temperatures on correlation functions is
particularly interesting for strongly interacting  multicomponent
systems (as has been emphasized  for the case of bose-bose and
fermi-fermi mixtures in \cite{CZS}), due to considerable change of the momentum distribution in the very narrow range of the
temperatures of the order of $E_f/\gamma.$ For the case of
bose-bose or fermi-fermi mixture it was possible\cite{CZS} to
obtain correlation functions only in the two limiting cases $T\ll
E_f/\gamma$ and  $E_f/\gamma\ll T \ll E_f.$ For bose-fermi
mixture, we are able to calculate correlation functions for any
ratio between $E_f/\gamma$ and $T \ll E_f $ (see fig.
\ref{nztfig}). By adding an imaginary part to $T,$ the procedure
presented in this section can be also easily generalized for non
equal time correlations.

\subsection{Low energy excitations in Tonks-Girardeau regime.}

As has been discussed in section \ref{numerics}, for $\gamma \gg
1$ there are two energy scales in the problem: the first energy
scale is the fermi energy of orbital motion $E_f=( \pi \hbar
n)^2/(2m),$ while the second is the the "spin wave" (relative
density oscillation) energy $E_f/\gamma.$ The second energy scale
is present only in strongly interacting multicomponent systems, as
has been emphasized earlier\cite{CZ,CZS}. Density profiles and correlation
functions we have considered earlier are
valid in the regime, when temperature is smaller than both of
these energy scales:
\begin{equation}
T\ll E_f/\gamma \ll E_f.\label{zeroT}
\end{equation}
However, interesting phenomena\cite{CZ,CZS,BalentsFiete, Matveev}
can be analyzed in the  "spin disordered"regime, when
\begin{equation}
E_f/\gamma\ll T \ll E_f.\label{spindisordered}
\end{equation}
This regime has attracted lots of attention recently in the
context of electrons in 1d quantum
wires\cite{CZ,BalentsFiete,Matveev}. In "spin disordered" regime,
"spin" degrees of freedom are completely disordered, while orbital
degrees are not affected much. From the point of view of orbital
degrees, this is still a low-temperature regime, since $T\ll E_f.$
The energy of the system doesn't change too much, while momentum
distribution changes dramatically as temperature changes from $0$
to the order of several $E_f/\gamma.$  "Spin disordered" regime
exists only for multicomponent systems and a crossover from true
ground state to "spin disordered" regime provides a unique
opportunity to study  the effects of low temperatures on a highly
correlated strongly interacting system. "Spin disordered" limit is
likely to be reached first in the experiments, and a significant
change of the density profile and of the momentum distribution as regime (\ref{zeroT}) is
reached can be used as a way to calibrate the temperatures much
smaller than $E_f.$

Only two limiting cases (\ref{zeroT})-(\ref{spindisordered}) have
been investigated for spin-$\frac12$ fermion and boson mixtures,
since in these cases "spin" wavefunctions are related to
eigenstates of  spin-$\frac12$ Heisenberg hamiltonian, and have a
complicated structure. In the case of bose-fermi mixture, "spin"
wavefunctions correspond to noninteracting fermionized single-spin
excitations, and one can calculate correlation functions in the
whole low-temperature limit, investigating crossover from true
ground state to "spin disordered" limit:

\begin{equation}
E_f/\gamma,T \ll E_f.\label{crossover}
\end{equation}

In the following calculations, we will neglect the influence of
nonzero temperature on orbital degrees, and will always assume
that orbital degrees are not excited. This assumption will affect
the results only at distances, at which the correlation functions
are already very small due to effects of spin excitations.

In the zeroth order in $1/\gamma$ expansion, energies of all spin
states are degenerate, and solutions of Bethe equations are given
by
\begin{eqnarray}
 \left(\frac{-\Lambda_\alpha+ic/2}{-\Lambda_\alpha-ic/2}\right)^N=1,\alpha=\{1,...,M\},\label{nonzeroTspectral1}\\
  e^{i k_j L}=\prod_{\beta=1}^{M}\frac{-\Lambda_\beta+ic/2}{-\Lambda_\beta-ic/2}, j=\{1, ... , N\}.
\label{nonzeroTspectral2}
\end{eqnarray}
In the next order in $1/\gamma$ expansion, both $k_j$ and
$\Lambda_i$ acquire corrections of the  order of $1/\gamma.$ Since
energy depends only on $\rho(k),$ we need to calculate corrections
to $\rho(k)$ in the leading order. According to
(\ref{bfequations31}), to calculate $1/\gamma$ correction to
$\rho(k),$ one can use $\Lambda_i$ in the zeroth order, given by
(\ref{nonzeroTspectral1}):
\begin{equation}
2\pi \rho(k)=1+\frac{1}{L}\sum_{i=1}^{M}\frac{4c}{4\Lambda_i^2+c^2}
\end{equation}
is independent of $k$ in the first order of $1/\gamma$ expansion.
If we define "spin" wave vectors according to
\begin{equation}
\frac{-\Lambda_\alpha+ic/2}{-\Lambda_\alpha-ic/2}=e^{i 2\pi
\kappa_\alpha/N},  \alpha=\{1,...,M\}, \label{nonzeroTspinrap}
\end{equation}
energy of the state with "spin" wave vectors $\kappa_i$ in
$1/\gamma$ order  is given by
\begin{equation}
E(\gamma, \kappa_i)=\frac{\pi^2}{3}\frac{N^2}{L^2}(N-\frac{4}{\gamma}\sum_{i=1}^M(1-\cos{\frac{2\pi \kappa_i}{N}})).
\label{egammakappa}
\end{equation}
Allowed values for "spin" wave vectors are
\begin{equation}
\hat K=\{\kappa_i \subset\{1,..., N\}, \kappa_i< \kappa_j \,\,
\mbox{for} \,\, i < j.\}\label{nonzeroTkappaset}
\end{equation}
The number of "spin" excitations (we will call
them magnons from now on) is fixed to be the number of
bosons, and different "spin" wave vectors cannot coincide.
Hence, magnons have a fermionic statistics. The
effect of nonzero temperatures is to average the correlations over
the different sets of possible $\kappa_i$ from
(\ref{nonzeroTkappaset}).

 According to (\ref{egammakappa}) in the first order
in $1/\gamma$ expansion magnons  do not interact with each other, and the
total energy is the sum of separate magnon energies.
 Magnon energy
spectrum is
\begin{equation}
\epsilon(\kappa)=\frac{4
\pi^2}{3\gamma}\frac{N^2}{L^2}(\cos{\frac{2\pi
\kappa}{N}}-1)=\frac{4E_f}{3\gamma}(\cos{\frac{2\pi
\kappa}{N}}-1). \label{magnonspectrum}
\end{equation}
Lowest state corresponds to $\kappa=N/2,$ and as the number of
magnons increases, "spin" wave vectors $\kappa$ near $N/2$ start
being occupied - (\ref{magnonspectrum}) proves the choice
(\ref{kappaset}) for the  true ground state at zero temperature.

\subsection{Density profiles}
\label{nztdp}
\begin{figure}
\psfig{file=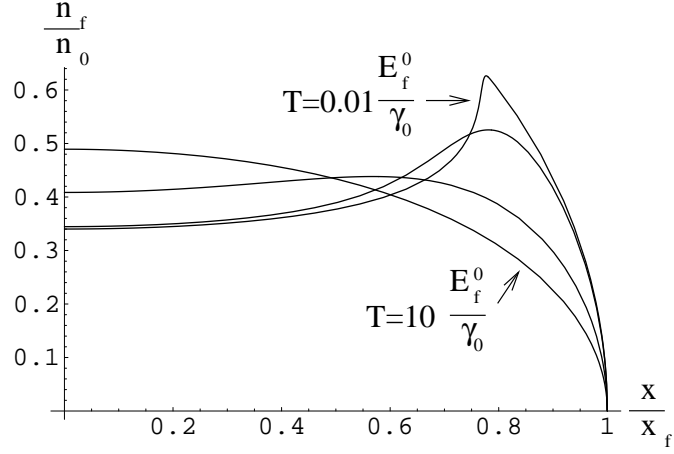} \caption{ \label{nztdensity} Evolution of the fermionic density
profile  in the strongly interacting regime as the function of temperature.
 Four graphs correspond to temperatures $0.01 E^0_f/\gamma_0,
0.2 E^0_f/\gamma_0, E^0_f/\gamma_0$ and $10 E^0_f/ \gamma_0.$
Here $\gamma_0$ is the Lieb-Liniger parameter in the center of the trap,
$x_f$ is the size of the fermionic trap, and $E^0_f=(\pi \hbar n_0)^2/(2m),$ where $n_0$
is the total density in the center of the trap.
 Overall number of bosons equals the number of fermions.
Non monotonous behavior of the fermi density profile persists up to $T\sim E^0_f/\gamma_0.$
The total density profile doesn't change considerably in this range of the temperatures, and is given by $n(x)=n^0 \sqrt{1-x^2/x_f^2}.$}
\end{figure}

In this subsection we will analyze the behavior of the strongly interacting mixture in a harmonic trap at low temperatures. Similarly to section \ref{LDAsection} we consider the case
\begin{equation}
\omega_b=\omega_f=\omega_0.
\end{equation}
According to (\ref{LDA2}), within the region of the coexistence densities are governed by
equations
\begin{eqnarray}
\mu^0_f(x)+ \frac{m \omega^2_0 x^2}2=\mu^0_f(0),\;
\mu^0_b(x)-\mu^0_f(x)=\mu^0_b(0)-\mu^0_f(0).\label{nztLDA}
\end{eqnarray}
Similarly to the case of $T=0,$ total density is given by (\ref{tgtotaldp}):
\begin{equation}
n^0(x)=n^0(0) \sqrt{1-\frac{x^2}{x_f^2}},
\label{nzttgtotaldp}
\end{equation}
and has a weak temperature dependence. On the other hand, relative density is controlled by solutions of the second equation (\ref{nztLDA}), and its dependence on temperature is quite
strong. It turns out, that in strongly interacting regime $\mu_b-\mu_f$ can be easily calculated
using formulas from the previous subsection. $\mu_b-\mu_f$ is the change of
the free energy, when one boson is added and one fermion is removed from the mixture.
On the language of the magnons this corresponds to an addition of one magnon. Therefore,
one obtains
\begin{equation}
\mu_b-\mu_f=\mu_m,
\end{equation}
where $\mu_m$ is the chemical potential of the magnons with energy spectrum (\ref{magnonspectrum}). As has been noted earlier, magnons obey fermionic statistics (only one magnon can occupy each state) and do not interact, so one can use Fermi distribution for their occupation number. Chemical potential for magnons $\mu_m$  as a function of $\alpha$ and $T$ can be obtained numerically from the normalization condition for the total number of magnons, which reads
\begin{equation}
\alpha=\int_0^{2\pi} \frac{1}{e^{\frac1{T}(\frac{4 E_f}{3\gamma }(\cos{k}-1)-\mu_m)}+1}\frac{dk}{2\pi}.
\label{nztnorm}
\end{equation}
After that, one can use LDA to obtain the density profiles. In fig. \ref{nztdensity}
we show the density of fermions for the case, when total number of bosons equals total number
of fermions. One sees, that density profile changes considerably at the temperatures of the order of $E^0_f/\gamma_0,$ where $E_f^0$ and $\gamma_0$ are the Fermi energy and Lieb-Liniger parameter in the center of the trap.
For $E^0_f/\gamma_0\ll T\ll E^0_f$ boson fraction $\alpha$ is uniform along the trap. As temperature is lowered, more bosons condense towards the center of the trap,
and fermionic density behaves non-monotonously as a function of the distance form the center of the trap.

\subsection{Fermi-Fermi correlations}
\label{fermiferminzt}
From now on we will consider the periodic boundary conditions, when the many body problem is
strictly solvable in the mathematical sense.  We will first describe the calculation of
fermi correlations, since it is simpler than calculation of  Bose
correlations. To calculate temperature averaged correlation
functions, we should be able to calculate
\begin{eqnarray}
\rho^f(0, \xi; T)=\frac{\sum_{\kappa_i \subset \hat K} e^{-\sum_i \epsilon(\kappa_i)/T}
(N-M) \int\Psi(\kappa_1, ...,  \kappa_M;x_1,x_2, ... ,0)\Psi^{\dagger}(\kappa_1, ...,  \kappa_M; x_1, x_2...,\xi) dx_1 ... dx_{N-1}}{\sum_{\kappa_i\subset \hat K} e^{-\sum _i \epsilon(\kappa_i)/T}}.
\label{nzTfermicorr}
\end{eqnarray}
Denominator in (\ref{nzTfermicorr}) is a partition function of noninteracting fermions in a micro canonical ensemble.
It can be written as
\begin{equation}
Z=\sum_{\kappa_i\subset \hat K} e^{-\sum _i \epsilon(\kappa_i)/T}=\int_0^{2\pi} e^{-i M \theta}\frac{d\theta}{2\pi} \prod_{\kappa=1}^N(1+e^{i \theta}e^{-\epsilon(\kappa)/T})
\label{nztZ}
\end{equation}
Numerator can be simplified using the factorization of "spin" and
orbital parts, similarly to (\ref{fermipsipsidaggerint}):
\begin{equation}
\rho^f(0, \xi; T)=\frac{1}Z  \frac{1}{(N-M-1)!M! L^N N^{M}}\sum_{d=1}^{N}\sum_{\kappa_i\subset \hat K}e^{-\sum_i \epsilon(\kappa_i)/T}
S^f(\kappa_1, ...,  \kappa_M;d) I(d,\xi;\kappa_1, ...,  \kappa_M).
\label{rhoft}
\end{equation}
Here $S^f(\kappa_1, ...,  \kappa_M;d)$ is an expression (\ref{sfd}) for an arbitrary choice of $\kappa_i$ belonging to (\ref{nonzeroTkappaset}):
\begin{eqnarray}
S^f(\kappa_1, ...,  \kappa_M;d)=\sum_{\hat y} \det[e^{i
\frac{2\pi}{N}\kappa_i y_j}]\det[e^{-i\frac{2\pi}{N} \kappa_i
y'_j}](-1)^y(-1)^{y'}. \label{nzTsfd}
\end{eqnarray}
$I(d,\xi;\kappa_1, ...,  \kappa_M)$ is an integral
(\ref{Idxidef}), which dependence on $\kappa_1, ...,  \kappa_M$
comes only through boundary conditions (\ref{nonzeroTspectral2}).
If $\sum_{i=1}^M \kappa_i \, \mbox{mod}\, N=D,$ where
$D=\{1,...,N-1\},$ then the set of $k_i$ which minimizes  kinetic
energy is uniquely defined:
\begin{equation}
\{ \frac{2\pi}{L}(-N/2+D/N),\frac{2\pi}{L}(-N/2+1+D/N), ... ,\frac{2\pi}{L}(N/2-1+D/N) \}.
\end{equation}
If $D=0,$ then there are two degenerate sets of $k_i,$ and each of them should be taken with a weight $1/2.$
Taking this into account, $I(d,\xi;\kappa_1, ...,  \kappa_M)$ can be expressed as
\begin{equation}
I(d,\xi;\kappa_1, ...,  \kappa_M)=I(d,\xi)
\sum_{D=0}^N(1-\frac{\delta_N(D)}2) \delta_N(D-\sum_{i=1}^M
\kappa_i) e^{-(D-\frac{N}2)\frac{2\pi i}{N}\xi},
\end{equation}
where
\begin{eqnarray}
\delta_N(x)=\left\{ \begin{array}{cc}
                              1 & \,\mbox{if} \, x \mbox{ mod } N = 0,\\
                  0 & \mbox{otherwise}
                              \end{array}
             \right.
\end{eqnarray}
$\delta_N(x)$ can be represented as a  Fourier sum,
\begin{equation}
\delta_N(x)=\frac1N \sum_{p=0}^{N-1}e^{\frac{2\pi i}N p x}.
\end{equation}
Taking this into account, correlation function (\ref{rhoft}) is
rewritten as
\begin{equation}
\rho^f(0, \xi; T)=\frac{1}Z  \frac{1}{(N-M-1)!M! L^N
N^{M+1}}\sum_{d=1}^{N}I(d,\xi) \sum_{D=0}^N(1-\frac{\delta_N(D)}2)
e^{-(D-\frac{N}2)\frac{2\pi i}{N}\xi}
\sum_{p=0}^{N-1}e^{\frac{2\pi i}N p D} S^f(d;p;T), \label{rhoft2}
\end{equation}
where
\begin{equation}
S^f(d;p;T)=\sum_{\kappa_i\subset \hat K}e^{-\sum_i(\frac{2\pi i}N p \kappa_i + \epsilon(\kappa_i)/T)}S^f(\kappa_1, ...,  \kappa_M;d) .
\end{equation}

 Calculation of $S^f(\kappa_1, ...,  \kappa_M;d)$ closely reminds a calculation of $S^f(d)$ in section \ref{ff}, so we will present only a brief derivation.
\begin{equation}
S^f(\kappa_1, ...,  \kappa_M;d)=\sum_{r=1}^{M+1} \frac{(N-M-1)!(M)!}{(r-1)!(M-r+1)!}S^f(\kappa_1, ...,  \kappa_M;d, r),
\end{equation}
 where $S^f(\kappa_1, ...,  \kappa_M;d, r)$ is a product of two determinants:
\begin{eqnarray}
S^f(\kappa_1, ...,  \kappa_M;d,r)=(-1)^{d-1} \sum_{P\subset S_{M}} \sum_{P'\subset S_{M}}(-1)^P (-1)^{P'}
\times \nonumber \\
 \prod_{i=1}^{r-1} ( \sum_{t_i=2}^{d} e^{i\frac{2\pi}{N} (\kappa_{P'_i}t_i-\kappa_{P_i} (t_i-1))})
  \prod_{i=r}^{M}(\sum_{t_i=d+1}^{N}  e^{i\frac{2\pi}{N} (\kappa_{P'_i}-\kappa_{P_i}) t_i}).
\end{eqnarray}
 From summation over $P, P'$ we can go to summation over $P, Q,$ where $P'= Q P.$ Also, one can analytically perform summation over $t_i$ in each
 of the brackets, since it is a  geometrical progression.
  \begin{eqnarray}
 S^f(\kappa_1, ...,  \kappa_M;d,r)=(-1)^{d-1} \sum_{P\subset S_{M}} \sum_{Q\subset S_{M}}(-1)^Q
 \prod_{i=1}^{r-1}g_f^1(d,\kappa_{Q_{P_i}},\kappa_{P_i})
 \prod_{i=r}^{M}g_f^2(d,\kappa_{Q_{P_i}},\kappa_{P_i}),
 \label{nztfermisdr}
 \end{eqnarray}
 where
 \begin{eqnarray}
 g^1_f(d, j, l)=e^{i\frac{2\pi}{N}l}\sum_{t=2}^{t=d}e^{i\frac{2\pi}{N}(j -l) t}\nonumber, \\
 g^2_f(d, j, l) = \sum_{t=d+1}^{t=N}e^{i\frac{2\pi}{N}(j-l) t}
 \end{eqnarray}
are independent of $r.$ We can use the "phase" variable integration trick to get rid of summation over $P,$ and then represent summation over $Q$ as a determinant:
 \begin{eqnarray}
 S^f(\kappa_1, ...,  \kappa_M; d, r)=(r-1)!(M-r+1)!(-1)^{d-1}\times  \nonumber \\
 \int_0^{2\pi} \frac{d\psi}{2\pi} e^{- i (r-1)\psi}
 \det\left[\begin{array}{cccc}
g^f (\psi,\kappa_1,\kappa_1)& g^f(\psi,\kappa_1,\kappa_2)&  ... &g^f(\psi,\kappa_1,\kappa_M)\\
  g^f(\psi,\kappa_2,\kappa_1) & g^f(\psi,\kappa_2,\kappa_2)& ... &g^f(\psi,\kappa_2,\kappa_M)\\
  ... & ... & ... & ...\\
  g^f(\psi,\kappa_M,\kappa_1) & g^f(\psi,\kappa_M,\kappa_2)& ... &g^f(\psi,\kappa_M,\kappa_M)
  \end{array}\right],
  \label{nztfermisdr2}
 \end{eqnarray}
  where
  \begin{equation}
  g^f(\psi,j,l)=e^{i\psi}g^1_f(d, j,l)+g^2_f(d, j,l).
  \label{nztgf}
  \end{equation}
  We can analytically perform summation over $r,$ since the form of the determinant and $g^f(\psi,j,l)$  are independent of $r,$ and combinatorial prefactor
  cancels in (\ref{nztfermisdr}). Similarly to (\ref{sdfinalfinal}) we represent summation over $r$ and integration over $\psi$ as a substitution $\psi_0=0,$  and obtain the following result:
  \begin{eqnarray}
  S^f(\kappa_1, ...,  \kappa_M;d)=(N-M-1)!M!(-1)^{d-1}
    \det\left[\begin{array}{cccc}
g^f (0,\kappa_1,\kappa_1)& g^f(0,\kappa_1,\kappa_2)&  ... &g^f(0,\kappa_1,\kappa_M)\\
  g^f(0,\kappa_2,\kappa_1) & g^f(0,\kappa_2,\kappa_2)& ... &g^f(0,\kappa_2,\kappa_M)\\
  ... & ... & ... & ...\\
  g^f(0,\kappa_M,\kappa_1) & g^f(0,\kappa_M,\kappa_2)& ... &g^f(0,\kappa_M,\kappa_M)
  \end{array}\right].
  \label{nztfermisdfinal}
  \end{eqnarray}
 To calculate $S^f(d;p;T)$ we have to sum (\ref{nztfermisdfinal}) for different choices of $\kappa_i$ with $\kappa_i$ dependent prefactor.
  One can take these prefactors into by multiplying each row in
 (\ref{nztfermisdfinal}) by
 \begin{equation}
 f(\kappa_i)=e^{-(\frac{2\pi i}N p \kappa_i+\epsilon(\kappa_i)/T)},
 \end{equation}
  since only one term from each row appears in the expansion of the determinant:
\begin{eqnarray}
S^f(d;p;T)=(N-M-1)!M!(-1)^{d-1}\times\nonumber\\
    \sum_{\kappa_1=1}^N ...\sum_{\kappa_M=1}^N
        \det\left[\begin{array}{cccc}
f(\kappa_1)g^f (0,\kappa_1,\kappa_1)& f(\kappa_1)g^f(0,\kappa_1,\kappa_2)&  ... &f(\kappa_1)g^f(0,\kappa_1,\kappa_M)\\
  f(\kappa_2)g^f(0,\kappa_2,\kappa_1) &f(\kappa_2)g^f(0,\kappa_2,\kappa_2)& ... &f(\kappa_2)g^f(0,\kappa_2,\kappa_M)\\
  ... & ... & ... & ...\\
  f(\kappa_M)g^f(0,\kappa_M,\kappa_1) &f(\kappa_M) g^f(0,\kappa_M,\kappa_2)& ... &f(\kappa_M)g^f(0,\kappa_M,\kappa_M)
  \end{array}\right].
  \label{nztfermisdtprefinal}
\end{eqnarray}
Summations over $\kappa_i$ in (\ref{nztfermisdtprefinal}) can be
performed analytically, since each choice of $\kappa_i$ is a term
in the expansion of the Fredholm determinant\cite{Smirnov}. The
desired expression has the form:
\begin{eqnarray}
S^f(d;p;T)=(N-M-1)!M!(-1)^{d-1}\int_0^{2\pi} \frac{d\theta}{2\pi}e^{-i (N-M)\theta}\times\nonumber\\
            \det\left[\begin{array}{cccc}
e^{i\theta}+f(1)g^f (0,1,1)& f(1)g^f(0,1,2)&  ... &f(1)g^f(0,1,N)\\
  f(2)g^f(0,2,1) &e^{i\theta}+f(2)g^f(0,2,2)& ... &f(2)g^f(0,2,N)\\
  ... & ... & ... & ...\\
  f(N)g^f(0,N,1) &f(N) g^f(0,N,2)& ... &e^{i\theta}+f(N)g^f(0,N,N)
  \end{array}\right].
  \label{nztfermisdtfinal}
  \end{eqnarray}
Integration over $\theta$ extracts terms from the determinant
which have $e^{i(N-M)\theta}$ dependence. Such terms appear, when
$N-M$ $e^{i\theta}$ elements in the expansion of the determinant
are taken along the diagonal. If $e^{i\theta}$ are chosen in the
rows except for $\kappa_1, ... , \kappa_M,$ then contribution from
such choice of $e^{i\theta}$ is a minor which equals
$f(\kappa_1)...f(\kappa_M)S^f(\kappa_1, ...,  \kappa_M;d).$ Thus
evaluation of the  prefactor in the $e^{i(N-M)\theta}$ dependence
of the determinant corresponds to summation  of
$f(\kappa_1)...f(\kappa_M)S^f(\kappa_1, ...,  \kappa_M;d)$ over
possible sets of $\kappa_i.$

Finally, substituting (\ref{nztfermisdtfinal}) into
(\ref{rhoft2}), one can evaluate numerically fermi-fermi
correlation functions for any temperature and ratio between boson
and fermion density in low temperature limit.

\begin{figure}
\psfig{file=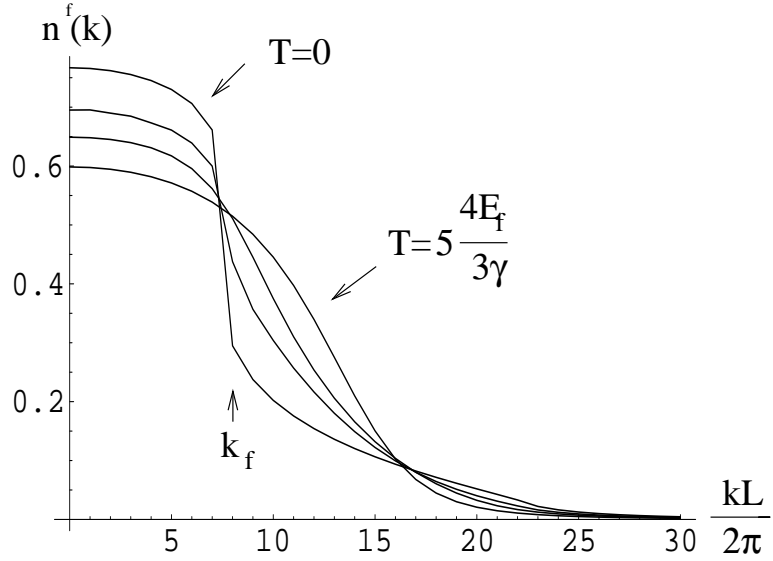} \caption{ \label{nztfig} Fourier  transform
of the Fermi-Fermi correlation function for $M=15, N=30.$ Four
graphs correspond to temperatures $0, 0.1\frac{4E_f}{3\gamma},
0.5\frac{4E_f}{3\gamma}, 5\frac{4E_f}{3\gamma}.$ In the range of
the temperatures $\sim E_f/\gamma \ll E_f$ fermi correlation
function changes considerably due to transition from true ground
state to "spin disordered" regime. In "spin disordered" fermi singularity at $k_f$ gets
completely smeared out by thermal "spin" excitations.}
\end{figure}

In fig. \ref{nztfig} we show numerically evaluated fermi-fermi
correlation function for $M=15 ,N=30$ and several temperatures,
ranging from $T=0$ to $T=5 \frac{4E_f}{3\gamma}.$ At this low
temperature region fermi-fermi correlation function changes
considerably due to transition from true ground state to "spin
disordered" regime. In "spin disordered" regime fermi singularity at $k_f$ gets completely
smeared out by thermal  "spin" excitations.

\subsection{Bose-Bose correlation function}
Bose-Bose correlation functions also change as $T$ goes up.
However,  since for $T=0$ $n^b(k)$ doesn't have any interesting
structure except for singularity at $k=0$, the effects of nonzero
temperatures will not be as dramatic as for fermi correlations. We
present here the results mainly for the sake of completeness.
Calculations in this subsection are similar to what has been done
in the previous subsection. Correlation function can be written as
\begin{eqnarray}
\rho^b(0, \xi; T)=\frac{\sum_{\kappa_i \subset \hat K} e^{-\sum_i \epsilon(\kappa_i)/T}
M \int\Psi(\kappa_1, ...,  \kappa_M; 0,x_2, ... ,x_N)\Psi^{\dagger}(\kappa_1, ...,  \kappa_M; \xi, x_2...,x_N) dx_2 ... dx_{N}}{\sum_{\kappa_i\subset \hat K} e^{-\sum _i \epsilon(\kappa_i)/T}}.
\label{nzTbosecorr}
\end{eqnarray}
Similarly to (\ref{rhoft2}), this can be written as
\begin{equation}
\rho^b(0, \xi; T)=\frac{1}Z  \frac{1}{(N-M)!(M-1)! L^N
N^{M+1}}\sum_{d=1}^{N}I(d,\xi) \sum_{D=0}^N(1-\frac{\delta_N(D)}2)
e^{-(D-\frac{N}2)\frac{2\pi i}{N}\xi}
\sum_{p=0}^{N-1}e^{\frac{2\pi i}N p D} S^b(d;p;T), \label{rhobt2}
\end{equation}
where
\begin{equation}
S^b(d;p;T)=\sum_{\kappa_i\subset \hat K}e^{-\sum_i(\frac{2\pi i}N p \kappa_i + \epsilon(\kappa_i)/T)}S^b(\kappa_1, ...,  \kappa_M;d) .
\end{equation}

Here $S^b(\kappa_1, ...,  \kappa_M;d)$ is an expression (\ref{sdprelim}) for an arbitrary choice of $\kappa_i$ belonging to (\ref{nonzeroTkappaset}):
\begin{eqnarray}
S^b(\kappa_1, ...,  \kappa_M;d)=\sum_{\hat y} \det[e^{i
\frac{2\pi}{N}\kappa_i y_j}]\det[e^{-i\frac{2\pi}{N} \kappa_i
y'_j}](-1)^y(-1)^{y'}. \label{nzTsbd}
\end{eqnarray}
Similarly to (\ref{sdrepr}), it can be written as
\begin{equation}
S^b(\kappa_1, ...,  \kappa_M;d)=\sum_{r=1}^{M} \frac{(N-M)!(M-1)!}{(r-1)!(M-r)!}S^b(\kappa_1, ...,  \kappa_M;d, r),
\label{nztsbrsum}
\end{equation}
where $S^b(\kappa_1, ...,  \kappa_M;d, r)$ is a result of the summation of (\ref{nzTsbd}) in the following subspace:
 \begin{equation}
\{1\leq y_2, ... , y_{r} \leq d <y_{r+1}, ... , y_{M}\leq N\}.
 \end{equation}
We can expand determinants of (\ref{nzTsbd}) using permutations:

\begin{eqnarray}
S^b(\kappa_1, ...,  \kappa_M;d,r)=(-1)^{d-1} \sum_{P\subset S_{M}} \sum_{P'\subset S_{M}}(-1)^P (-1)^{P'}
\times \nonumber \\
 e^{i\frac{2\pi}{N} (\kappa_{P'_1}1-\kappa_{P_1}d)}\prod_{i=2}^{r} ( \sum_{y_i=2}^{d} e^{i\frac{2\pi}{N} (\kappa_{P'_i}y_i-\kappa_{P_i} (y_i-1))})
  \prod_{i=r+1}^{M}(\sum_{y_i=d+1}^{N}  e^{i\frac{2\pi}{N} (\kappa_{P'_i}-\kappa_{P_i})y_i}).
\end{eqnarray}
 From summation over $P, P'$ we can go to summation over $P, Q,$ where $P'= Q P.$ Also , one can analytically perform summation over $y_i$ in each
 of the brackets, since it is a  geometrical progression. Compared to the case of fermions, there are 3 types of the brackets:

 \begin{eqnarray}
 S^f(\kappa_1, ...,  \kappa_M;d,r)=(-1)^{d-1} \sum_{P\subset S_{M}} \sum_{Q\subset S_{M}}(-1)^Q
 g_b^0(d,\kappa_{Q_{P_1}},\kappa_{P_1})
 \prod_{i=2}^{r}g_b^1(d,\kappa_{Q_{P_i}},\kappa_{P_i})
 \prod_{i=r}^{M}g_b^2(d,\kappa_{Q_{P_i}},\kappa_{P_i}),
 \label{nztbosesdr}
 \end{eqnarray}
 where
 \begin{eqnarray}
 g^0_b(d, j, l)=e^{i\frac{2\pi}{N}(j 1 - l d)}\nonumber, \\
 g^1_b(d, j, l)=e^{i\frac{2\pi}{N}l}\sum_{t=2}^{t=d}e^{i\frac{2\pi}{N}(j -l) t}\nonumber, \\
 g^2_b(d, j, l) = \sum_{t=d+1}^{t=N}e^{i\frac{2\pi}{N}(j-l) t}
 \end{eqnarray}
 We can use "phase integration" trick to represent (\ref{nztbosesdr}) as an integral of some determinant, but there will be two phase variables,
 since there are 3 types of inequivalent brackets:
  \begin{eqnarray}
 S^b(\kappa_1, ...,  \kappa_M; d, r)=(r-1)!(M-r)!(-1)^{d-1}\times  \nonumber \\
 \int_0^{2\pi} e^{- i (r-1)\psi} \frac{d\psi}{2\pi}\int_0^{2\pi} \frac{d\phi}{2\pi} e^{-i \phi}
 \det\left[\begin{array}{cccc}
g^b (\psi,\phi,\kappa_1,\kappa_1)& g^b(\psi,\phi,\kappa_1,\kappa_2)&  ... &g^b(\psi,\phi,\kappa_1,\kappa_M)\\
  g^b(\psi,\phi,\kappa_2,\kappa_1) & g^b(\psi,\phi,\kappa_2,\kappa_2)& ... &g^b(\psi,\phi,\kappa_2,\kappa_M)\\
  ... & ... & ... & ...\\
  g^b(\psi,\phi,\kappa_M,\kappa_1) & g^b(\psi,\phi,\kappa_M,\kappa_2)& ... &g^b(\psi,\phi,\kappa_M,\kappa_M)
  \end{array}\right],
  \label{nztbosesdr2}
 \end{eqnarray}
where
\begin{equation}
g^b(\psi,\phi,j,l)=e^{i\phi}g^0_b(d, j,l)+e^{i\psi}g^1_b(d, j,l)+g^2_b(d, j,l).
\label{nztgb}
\end{equation}
After integration over $\phi,$ determinant in (\ref{nztbosesdr2})
has terms up to  $e^{i(M-1)\psi},$ therefore integration over
$\psi$ and summation according to (\ref{nztsbrsum}) are
equivalent to substitution $\psi_0=0:$
  \begin{eqnarray}
  S^b(\kappa_1, ...,  \kappa_M;d)=(N-M)!(M-1)!(-1)^{d-1}
    \int_0^{2\pi} \frac{d\phi}{2\pi} e^{-i \phi}\times \nonumber \\ \det\left[\begin{array}{cccc}
g^b (0,\phi,\kappa_1,\kappa_1)& g^b(0,\phi,\kappa_1,\kappa_2)&  ... &g^b(0,\phi,\kappa_1,\kappa_M)\\
  g^b(0,\phi,\kappa_2,\kappa_1) & g^b(0,\phi,\kappa_2,\kappa_2)& ... &g^b(0,\phi,\kappa_2,\kappa_M)\\
  ... & ... & ... & ...\\
  g^b(0,\phi,\kappa_M,\kappa_1) & g^b(0,\phi,\kappa_M,\kappa_2)& ... &g^b(0,\phi,\kappa_M,\kappa_M)
  \end{array}\right].
  \label{nztbosesdfinal}
  \end{eqnarray}
Integral over $\phi$ can be simplified further, since the
determinant in (\ref{nztbosesdfinal}) has a form $A_0+A_1
e^{i\phi}.$ The form above follows from the fact that a part of
the matrix which depends on $e^{i\phi}$ has a rank 1 and the
formula for the determinant of the sum of the matrices(see page
221 of \cite{KBI}). Let's for a moment introduce a notation
$z=e^{i\phi}.$ Integration over $\phi$ with a weigh
$e^{-i\varphi}$extracts the term $A_1,$ which can be alternatively
written as a difference between two determinants, one when $z=1$
and the other when $z=0$ ($g^f(\psi,j,l)$ is given by
(\ref{nztgf}))
\begin{eqnarray}
  S^b(\kappa_1, ...,  \kappa_M;d)=(N-M)!(M-1)!(-1)^{d-1}
    \det\left[\begin{array}{cccc}
g^b (0,0,\kappa_1,\kappa_1)& g^b(0,0,\kappa_1,\kappa_2)&  ... &g^b(0,0,\kappa_1,\kappa_M)\\
  g^b(0,0,\kappa_2,\kappa_1) & g^b(0,0,\kappa_2,\kappa_2)& ... &g^b(0,0,\kappa_2,\kappa_M)\\
  ... & ... & ... & ...\\
  g^b(0,0,\kappa_M,\kappa_1) & g^b(0,0,\kappa_M,\kappa_2)& ... &g^b(0,0,\kappa_M,\kappa_M)
  \end{array}\right]-\nonumber\\
  (N-M)!(M-1)!(-1)^{d-1}\det\left[\begin{array}{cccc}
g^f (0,\kappa_1,\kappa_1)& g^f(0,\kappa_1,\kappa_2)&  ... &g^f(0,\kappa_1,\kappa_M)\\
  g^f(0,\kappa_2,\kappa_1) & g^f(0,\kappa_2,\kappa_2)& ... &g^f(0,\kappa_2,\kappa_M)\\
  ... & ... & ... & ...\\
  g^f(0,\kappa_M,\kappa_1) & g^f(0,\kappa_M,\kappa_2)& ... &g^f(0,\kappa_M,\kappa_M)
  \end{array}\right].
  \label{nztbosesdfinalfinal}
  \end{eqnarray}

  We note, that a similar trick is explained on the page 609 of \cite{IzerginPronko}. After that, summation over different $\kappa_i$ can be performed
similarly to the case of fermions:
\begin{eqnarray}
S^b(d;p;T)=(N-M)!(M-1)!(-1)^{d-1}\int_0^{2\pi} \frac{d\theta}{2\pi}e^{-i (N-M)\theta}\times\nonumber\\
            \det\left[\begin{array}{cccc}
e^{i\theta}+f(1)g^b (0,0,1,1)& f(1)g^b(0,0,1,2)&  ... &f(1)g^b(0,0,1,N)\\
  f(2)g^b(0,0,2,1) &e^{i\theta}+f(2)g^b(0,0,2,2)& ... &f(2)g^b(0,0,2,N)\\
  ... & ... & ... & ...\\
  f(N)g^b(0,0,N,1) &f(N) g^b(0,0,N,2)& ... &e^{i\theta}+f(N)g^b(0,0,N,N)
  \end{array}\right] -\frac{N-M}{M}S^f(d;p;T),
  \label{nztbosesdtfinal}
  \end{eqnarray}
where $S^f(d;p;T)$ is defined in (\ref{nztfermisdtfinal}).

\section{Experimental considerations and conclusions}
\label{experiments}

In this section we will consider in detail possible ways to realize the system under investigation in experiments with cold atoms.

An array of one dimensional tubes of cold atoms along $x$ direction has been realized experimentally using strong optical lattices in two dimensions\cite{Weiss,Paredes,Moritz1dmolecules,Moritz,Greiner1d} $y$ and  $z$.
The large number of tubes provides a good imaging quality, but the number of atoms  and the ratio between bose and fermi particle numbers varies from tube to tube, and may complicate
the interpretation of the experiments (one of the ways to fix the ratio between bose and fermi numbers for all tubes will be discussed later). In addition, due to harmonic confinement along the axis of the tube, bose and fermi densities vary within each tube, which causes non-homogeneous broadening of the momentum distribution.
Alternatively, single copies of one dimensional mixtures with  constant densities along
the axis  can be realized in micro traps on a chip\cite{chips}, or using cold atoms in a 1d box
potential\cite{BECinbox}. Here we will mostly concentrate on a realization of 1d system using
strong 2D optical lattice in $y$ and $z$ directions.

First of the conditions (\ref{intercond}), $m_b=m_f,$ is approximately satisfied for isotopes of the atoms, and one can expect our theory to be valid with high accuracy for them. Some of the promising candidates are
$^{39(41)}K-^{40}K$\cite{Cote}, $^{171}Yb+^{172}Yb$\cite{Yb},  and
$^{86(84)}Rb-^{87(85)}Rb$\cite{Rb}. Different isotopes of potassium
have already been cooled to quantum degeneracy \cite{bfexp,K41BEC} by sympathetic
cooling with $Rb.$ There is another way to satisfy
the first condition of (\ref{intercond})
using already available degenerate mixtures\cite{bfexp}.
If one uses  an additional optical lattice along  the $x$ direction with filling
factors much smaller than one, then  (\ref{initialhamiltoniansc}) is an
effective Hamiltonian describing this system with the effective masses
determined by the tunneling, similarly to a recent realization of Tonks-Girardeau gas for bosons\cite{Paredes}. Finally, we note that one can realize experimentally the model, which has the same energy eigenvalues as (\ref{initialhamiltoniansc}), using a mixture of two {\it bosonic} atoms (see next paragraph). If one chooses two magnetic sublevels of the same atom,
equality of masses will be satisfied automatically.

Second of the conditions (\ref{intercond}), $g_{bb}=g_{bf}>0,$ can also be satisfied in current experiments, using a combination of several approaches. First, one can use Feshbach resonances to control the interactions: this is particularly straightforward for $Li-Na$ of $K-Rb$ mixtures, where resonances have already been observed experimentally\cite{LiNaFeshbach,KRbFeshbach}.
Second, we point out that it is sufficient to have equal (positive) signs
for the two scattering lengths, but not necessarily their
magnitudes. Well away from confinement induced
resonances\cite{Olshanii98}, 1D interactions are given by $
g_{bb}=2\hbar \omega_{b\perp} a_{bb}, g_{bf}=2\hbar
\sqrt{\omega_{b\perp}\omega_{f\perp}}a_{bf}, $ where
$\omega_{b\perp},\omega_{f\perp}$ are radial confinement
frequencies, and $a_{bb},a_{bf}$ are 3D scattering lengths. For a
fixed value of $a_{bb}/a_{bf},$ one can always choose the detuning
of the  optical lattice laser frequencies in such a way  that $g_{bb}=g_{bf}.$
After that, one can vary the intensity of the $y,z$ optical lattice beams and change $g$, while always being on the integrable line of the phase diagram.
Combination of these two approaches to control 1D interactions gives a lot of freedom
for experimental realization of equal one dimensional interactions. Finally, lets describe how to realize the bosonic model, which has the same eigenvalues as the model (\ref{initialhamiltoniansc}). Bosonic system is characterized by $3$ interaction parameters,
$g_{11}, g_{22}, g_{12}.$ If one tunes $g_{11}$  to $+\infty,$ then bosons of type $1$ get "fermionized" within the same type, and the model will be equivalent in terms of energy spectrum, density profiles and collective modes to (\ref{initialhamiltoniansc}). Note, however, that single-particle correlation functions will be different, and the results of  sections \ref{zerotcorrfunc} and \ref{nztcorrelations} (except for \ref{nztdp}) are not applicable. This  general equivalence between bose-bose and bose-fermi models is valid for any ratio between $g_{22}$ and $g_{12}.$ One can push this result even further, by tuning $g_{22}$ to $+\infty.$ In this case eigenstates of (\ref{initialhamiltoniansc}) are equivalent to spin$-1/2$ fermi system\cite{Yang67, Gaudin, Recati}, and some predictions
for those systems can be applied for bosons.

Detection of the properties of the system may be hindered by the fact, that
both number of atoms and relative fraction of bosons $\alpha$ vary from tube to tube.
However, one can use Feshbach resonances to fix the boson fraction  to be $\alpha=1/2$ in each tube \footnote{We thank G. Modugno for pointing out this possibility.}. To do this, one can use Feshbach resonance for bose-fermi scattering to adiabatically create molecules before loading the mixture in strong $y,z$ optical lattice. If one gets rid of unpaired atoms at this stage, switches on $y,z$ optical lattice, and adiabatically dissociates the molecules,
boson fraction will be fixed in each tube to be $\alpha=1/2.$ Most of our figures have
been calculated for this particular boson fraction. Our results in harmonic traps are presented as functions of $\gamma_0=m g/(\hbar^2n_0),$ where $n_0$ is a total
density in the center of a one dimensional trap, and $\gamma_0$ is the Lieb-Liniger \cite{LL} parameter in the center of the trap. $\gamma_0\gg1$ corresponds to a strongly interacting regime. $n_0$ varies from tube to tube,
and to be able to compare theoretical predictions precisely with experiments, one should
be able to have an optical access to regions where variation of $n_0$ is small.

Most of our experimental predictions, except for those in section \ref{nztcorrelations},
deal with zero temperature case. Experimentally, one needs to verify the quantum degeneracy
of the gases in 1D regime. A possible way to identify the onset of quantum degeneracy is based on density profiles\cite{fermipressure}. In Figs. \ref{meanfigure} and \ref{lgdp} we show the
density profiles at zero temperature for weak and strong interactions, when the harmonic confinement frequency $\omega_0$ is the same for bosons and fermions. In both cases, only central part is occupied by bosons, and outer shells consist of fermions only. In addition, for the strong
interactions fermi density develops a strong peak at the edge of bosonic cloud.
When the interactions are not strong ($\gamma_0\lesssim 1$), one can estimate the temperature at which quantum effects become important for ground state density profile to be of the order of $N\hbar \omega_0,$ where $N$ is the total number of atoms in a tube.
In the strongly interacting regime ($\gamma_0\gg 1$), however, situation is very different. There are two temperature scales in the problem: $E^0_f=(\pi \hbar n_0)^2/(2m),$ and $E^0_f/\gamma_0 \ll E^0_f.$ As the temperature goes up from $0$ to $\sim E^0_f/\gamma_0,$ density profile changes as shown in figure \ref{nztdensity}, and the peak in the fermion density disappears.  However, total density profile doesn't change much as long as $T\ll E^0_f.$ This effect can be qualitatively understood as the demonstration of the "fermionization" of the bose-fermi cloud, as will be explained in the next paragraph.

First, lets consider the case without a harmonic potential. When interactions are strong,
bosons tend to avoid fermions and other bosons. Whenever coordinates of any two particles
coincide, wavefunction is close to $0.$ Effectively, the gas is mutually "fermionized", and the ground state energy of the system is close to the ground state energy of the pure  noninteracting fermi gas with a density equal to the {\it total} density of bosons and
fermions. Dependence of the energy on the relative density (or boson fraction $\alpha$) appears only in the next order in $1/\gamma$ expansion, and two first terms in this expansion are given by (\ref{largegammaenergy}). Since dependence of the energy on boson
fraction $\alpha$ is $\gamma \gg 1$ times smaller than dependence on total density, the "quantum degeneracy" temperature for relative density excitations is also $\gamma$ times smaller
than quantum degeneracy temperature for fermions with density $n,$ hence it is $\sim E_f/\gamma.$ When harmonic trap is present at $T=0,$ relative density distributes itself
to minimize the total energy. As temperature becomes of the order of several $ E^0_f/\gamma_0,$ almost all relative density modes get excited, and boson fraction becomes uniform along the trap.
Total density modes are still not excited, since their quantum degeneracy temperature is
$E^0_f,$ and therefore the total density profile doesn't change much. Temperature $E^0_f/\gamma_0,$ is important not only for density distribution, but also for correlation
functions, as will be discussed later.

Knowledge of the exact dependence of the energy as the function of densities and interactions
allows to investigate not only the static properties, but also dynamic behavior. In section \ref{LDAsection} we developed a two-fluid hydrodynamic approach to calculate the frequencies
of collective oscillations. In the strongly interacting limit we predict the appearance of
low-lying modes, with a frequency scaling as $\sim \omega_0/\sqrt{\gamma_0}.$  These modes
correspond to "out of phase" oscillations of bose and fermi clouds that keep the total density approximately constant. These modes can be understood as follows: due to fermionization effects discussed in previous paragraph, for $\gamma_0 \gg 1$
the energetic penalty for changing the relative density of bosons and fermions is small, and hence it doesn't cost too much energy to create "out of phase" oscillations that don't change the total density. Dependence of the frequencies of low-lying oscillations with small quantum numbers  on overall boson fraction in a tube is shown in figure \ref{tgmodes}.
In addition to  low lying "out of phase" oscillations, the cloud has "in phase" oscillations,
with the frequencies $\omega_n=n \omega_0,$ similarly to Tonks-Girardeau gas of bosons\cite{Menotti}. These modes have frequencies considerably higher than "out of phase" modes, and are not shown in figure \ref{tgmodes}.
 One can excite any of these excitations by adding a perturbation
of the matching frequency, similarly to what has been done to bosons in \cite{Moritz}.
A different manifestation of the slow "out of phase" dynamics can be observed looking at the evolution of density perturbations: initial perturbation will split into fast "in phase" part,
moving at fermi velocity, and slow "out of phase" part. This is similar to "spin-charge separation", proposed for fermi\cite{Recati} or bose\cite{Fuchs}  spin $1/2$ mixtures.
When interactions are not strong ($\gamma_0 \lesssim 1$), one can obtain frequencies of all modes using mean-field energy. Figure \ref{mfmodes} shows the dependence of frequencies
for equal number of bosons and fermions ($\alpha=1/2$) on $\gamma_0.$ Even in mean field regime frequency of "out of phase" oscillations gets smaller as interactions get stronger.
Already for $\gamma_0\approx 1$ results for $\gamma_0\gg1$ extrapolate mean-field results very well.

\begin{figure}
\psfig{file=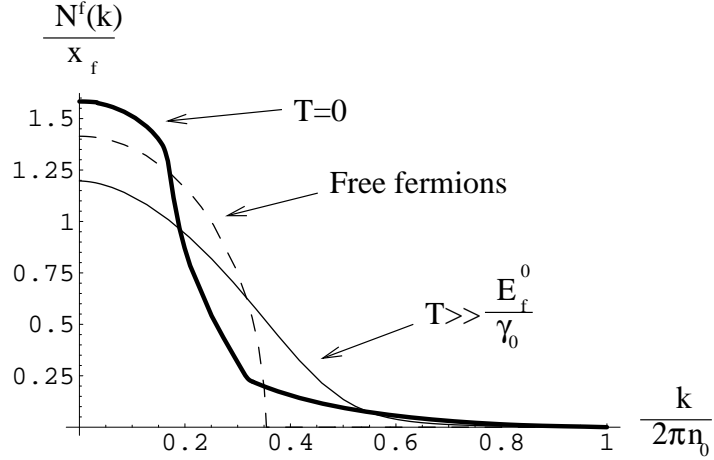} \caption{ \label{nztldacorr} Momentum distribution
 for fermions after averaging by inhomogeneous density profile for harmonic confinement. Results are shown for $T=0,$ (thick line), $E^0_f/\gamma_0 \ll T \ll E^0_f$ (normal line) and for  the same number of noninteracting fermions (dashed). Overall number of bosons in a trap equals the
 total number of fermions, $n_0$ is the total density in the center of the trap, $E^0_f=(\pi \hbar n_0)^2/(2m), \; \gamma_0 \gg 1$ is the Lieb-Liniger parameter in the center of the trap, and $2x_f$ is the total size of the cloud. In the range of
the temperatures $\sim E^0_f/\gamma_0 $ fermi correlation
function changes considerably due to transition from true ground
state to "spin disordered" regime. }
\end{figure}

Finally, lets discuss theoretically the most interesting and sensitive measure of the correlations, single particle correlation function, considered in sections \ref{zerotcorrfunc} and \ref{nztcorrelations}. Fourier transform of the single particle correlation function is an occupation number, and it can be measured experimentally using Bragg spectroscopy\cite{Bragg} or time of flight
measurements\cite{Paredes}. We can calculate these correlation
functions in strongly interacting regime under periodic boundary conditions for any temperatures. At zero temperature bose momentum distribution has a singularity  (\ref{nbksingularity}) at
$k=0$ reminiscent of BEC in higher dimensions, and its strength is controlled by Luttinger
liquid parameter $K_b,$ which depends only on boson fraction for strong interactions.
 For fermions, momentum distribution has a lot of interesting features. At zero temperature, several momentum distributions are presented in figs.
\ref{fermink}, \ref{fermink1}, \ref{fermink2}. One sees, that due to strong interactions,
fermi step at $k_f$ gets smeared out even at $T=0,$ and $n^f(k)$ is considerably different from $0$ at wave vectors far away from $k_f.$ However, total change of $n^f(k)$ as one crosses $k_f$ is quite large. In addition, $n^f(k)$ develops an extra singularity\cite{Frahm} at $k_f+2k_b,$ and the strength of this singularity is higher
for small boson fractions. As the temperature rises, momentum distribution changes considerably in the region of low temperatures of the order of $E_f/\gamma,$ and
its evolution as a function of temperature is shown in figure \ref{nztfig}. For $E_f/\gamma \ll T \ll E_f,$ one enters so called "spin disordered" regime\cite{CZ,CZS},
where singularity at $k_f$ gets completely washed out, and for equal densities of bosons and fermions momentum distribution  gets almost twice as wide compared to $T\ll E^0_f/\gamma$. A strong change of the
momentum distribution in a small range of temperatures can be used to perform a thermometry
at very small temperatures. To verify experimentally exact numerical correlation functions one needs to work with systems at constant densities along $x$ direction. Such constant density can be achieved in  experiments with micro traps\cite{chips},
or in 2D arrays of tubes, if one makes a very shallow harmonic confinement, and creates
strong box-like impenetrable potential at the sides of the tubes with the help of additional lasers. If the system is in harmonic trap, lots of the features of correlations themselves (i.e. singularity at $k_f+2k_b$) get washed out due to averaging over inhomogeneous density profile\cite{Gerbier}. However, the averaged correlation function still shows significant change in the region of temperatures of the order of $E^0_f/\gamma_0,$ and the results for $T\ll E^0_f/\gamma_0$ and $E^0_f/\gamma_0 \ll T \ll E^0_f$  are shown in fig. \ref{nztldacorr}. The point where $N^f(k)$ has a discontinuous derivative for $T=0$ corresponds to the fermi wavevector for the maximal density of fermions (at the edge of the bosonic cloud). For comparison, we also show $N^f(k)$ for the same number of fermions in the same trap for noninteracting case.

In conclusion, we presented a model for interacting bose-fermi mixture in 1D, which is  exactly solvable by Bethe ansatz technique. We obtained the energy numerically in the thermodynamic limit, and used it to prove the absence of the demixing under conditions (\ref{intercond}), contrary to  prediction of a mean field approximation. Combining exact solution with local density approximation (LDA) in a harmonic trap, we calculated the density profiles and frequencies of collective modes in various limits. In the strongly interacting regime, we predicted the appearance of low-lying
collective oscillations which correspond to the counterflow of the two species. In the strongly interacting regime we used exact wavefunction to calculate the single particle correlation functions for bosons and fermions at zero temperature under periodic boundary conditions. We derived an analytical formula, which allows to calculate correlation functions at all distances numerically for a polynomial time in system size. We investigated
numerically two strong singularities of the momentum distribution for fermions at $k_f$ and $k_f+2k_b.$ We extended the results for correlation functions for low temperatures, and calculated correlation functions in the crossover regime from $T=0$ to "spin disordered" regime. We also calculated the evolution of the density profile in a harmonic trap
at small nonzero temperatures. We showed, that in strongly interacting regime correlation functions change dramatically as temperature changes from $0$ to a small temperature $\sim E_f/\gamma \ll E_f,$ where $E_f=(\pi \hbar n)^2/(2m), \; n$ is the total density and $\gamma$ is the Lieb-Liniger parameter. Finally, we analyzed the experimental situation, proposed several ways to implement the exactly solvable hamiltonian and combined the results for correlation functions with LDA.

We thank M. Lukin, L. Mathey, G. Shlyapnikov, D.Petrov, P.Wiegmann, C. Menotti and D.W. Wang for useful discussions. This work was partially supported by the NSF
grant DMR-0132874.

\appendix
\section{}
\label{appA}
 In this appendix we will prove that all solutions of
equations (\ref{bfequation1})-(\ref{bfequation2})
\begin{eqnarray}
  \prod_{i=1}^{N}\frac{k_i-\Lambda_\alpha+ic/2}{k_i-\Lambda_\alpha-ic/2}=1,\alpha=\{1,...,M\},\label{app1}\\
  e^{i k_j L}=\prod_{\beta=1}^{M}\frac{k_j-\Lambda_\beta+ic/2}{k_j-\Lambda_\beta-ic/2}, j=\{1, ... , N\}
\label{app2}
\end{eqnarray}
are always real. This is a major simplification  for the analysis
of the excited states compared to spin-$\frac12$ fermion systems,
where one has to consider complex solutions\cite{babooks}.

Suppose that solutions of (\ref{app1})-(\ref{app2}) are complex
numbers, such that
\begin{eqnarray}
\inf \mbox{Im}\, k_j=k^-\leq\sup \mbox{Im}\,k_j=k^+,\\ \inf
\mbox{Im}\,\Lambda_\alpha=\Lambda^-\leq\sup
\mbox{Im}\,\Lambda_\alpha=\Lambda^+.
\end{eqnarray}
We need to prove that $k^-=k^+=\Lambda^-=\Lambda^+=0.$

First, lets prove that
\begin{eqnarray}
k^-\leq\Lambda^-, \label{app3} \\ \Lambda^+\leq k^+.\label{app4}
\end{eqnarray}
 Suppose that (\ref{app3}) is not valid, i. e.
\begin{equation}
\exists \; \alpha : \mbox{Im}\,k_j - \mbox{Im}\,\Lambda_\alpha >0
\;\forall \; j.
\end{equation}
Then
\begin{equation}
\left|\frac{k_j-\Lambda_\alpha+ic/2}{k_j-\Lambda_\alpha+ic/2}
\right|>1  \; \forall \;  j,
\end{equation}
and absolute value of the lhs of eq. (\ref{app1}) is bigger than
1, which contradicts the equation. Equation (\ref{app4}) can be
proven similarly.

Now, lets prove that
\begin{eqnarray}
k^+\leq0,\label{app5}\\ k^-\geq0. \label{app6}
\end{eqnarray}
These equations together with (\ref{app3})-(\ref{app4}) would
imply $k^-=k^+=\Lambda^-=\Lambda^+=0.$

Suppose that (\ref{app5}) is not valid, i. e. $\exists \; j :
\mbox{Im}\,k_j=k^+ >0.$ From (\ref{app4}) it follows that
\begin{equation}
\mbox{Im}\,k_j - \mbox{Im}\,\Lambda_\beta \geq 0 \;\forall \;
\beta,
\end{equation}
therefore
\begin{equation}
\left|\frac{k_j-\Lambda_\beta+ic/2}{k_j-\Lambda_\beta+ic/2}
\right| \geq 1 \; \forall \;  \beta,
\end{equation}
 and absolute value of the rhs of equation (\ref{app2}) is not
 smaller than 1. On the other hand, by assumption lhs of this equation is
 smaller than 1 :
 \begin{equation}
 |e^{ik_jL}|=e^{-k^+ L}<1.
 \end{equation}
Contradiction proves the validity of (\ref{app5}), and
(\ref{app6}) can be proven similarly.


\begin{thebibliography}{10}
\bibitem{bfexp}B. DeMarco and D.S. Jin, Science, {\bf 285}, 1703(1999);
 F. Schreck {\it et al.}, Phys. Rev. Lett. {\bf 87}, 080403 (2001);
G. Modugno {\it et al.}, Science  {\bf 297}, 2240 (2002); Z. Hadzibabic {\it et al.},
 Phys. Rev. Lett. {\bf 88}, 160401 (2002);
 G. Roati  {\it et al.}, Phys. Rev. Lett. {\bf 89}, 150403 (2002);
  J. Goldwin {\it et al.},  Phys. Rev. A {\bf 70}, 021601(R) (2004).
\bibitem{fermipressure}A.G. Truscott {\it et al.}, Science {\bf 291}, 2570(2001).
\bibitem{KRbFeshbach}A. Simoni {\it et al.}, Phys. Rev. Lett. {\bf 90}, 163202 (2003);
S. Inouye {\it et al.}, Phys. Rev. Lett. {\bf 93}, 183201 (2004);
F. Ferlaino {\it et al.}, cond-mat/0510630.
\bibitem{LiNaFeshbach} C.A. Stan {\it et al.}, Phys. Rev. Lett. {\bf 93}, 143001 (2004).
\bibitem{Jaksch98}D. Jaksch {\it et al.}, Phys. Rev. Lett. {\bf 81}, 3108 (1998).
\bibitem{Bloch} M. Greiner {\it et al.}, Nature {\bf 415}, 39(2002).
\bibitem{Weiss} T. Kinoshita,  T. Wenger and D.S. Weiss,  Science, {\bf 305}, 1125 (2004).
\bibitem{Paredes} B. Paredes {\it et al.}, Nature {\bf 429}, 277 (2004).
\bibitem{Moritz1dmolecules} H. Moritz {\it et al.}, Phys. Rev. Lett. {\bf 94}, 210401 (2005).
\bibitem{bfshort} A. Imambekov and E.Demler, cond-mat/0505632.
\bibitem{bftheory}K. Molmer, Phys. Rev. Lett. {\bf 80}, 1804 (1998);
 L. Viverit, C. J. Pethick and  H. Smith, Phys. Rev.A {\bf 61}, 053605 (2000);
   H. Heiselberg {\it et al.}, Phys. Rev. Lett. {\bf 85}, 2418 (2000);
M. J. Bijlsma, B. A. Heringa, and H. T. C. Stoof, Phys. Rev. A {\bf 61}, 053601 (2000);
 L. Viverit and S. Giorgini , Phys. Rev. A  {\bf 66}, 063604 (2002);
 A. Albus, F. Illuminati and J. Eisert, Phys. Rev. A {\bf 68}, 023606 (2003);
 H. P. Buchler and G. Blatter, Phys. Rev. Lett. {\bf 91}, 130404 (2003);
 M. Lewenstein {\it et al.}, Phys. Rev. Lett. {\bf 92}, 050401 (2004);
 D.-W. Wang, M.Lukin and E.Demler, cond-mat/0410494;
 A. Storozhenko {\it et al.}, Phys. Rev. A {\bf 71},063617 (2005).
\bibitem{Das}K.K. Das,  Phys. Rev. Lett. {\bf 90}, 170403 (2003).
\bibitem{CazalillaHo}M. A. Cazalilla and  A. F. Ho, Phys. Rev. Lett. {\bf 91}, 150403 (2003).
\bibitem{jap_numerics} Y. Takeuchi and H. Mori, cond-mat/0508247;cond-mat/0509048;cond-mat/0509393.
\bibitem{Mathey} L. Mathey {\it et al.}, Phys. Rev. Lett. {\bf 93}, 120404 (2004).
\bibitem{CDW}T. Miyakawa, H. Yabu and  T. Suzuki, Phys. Rev. A {\bf 70}, 013612 (2004);
E. Nakano and H.Yabu, Phys. Rev. A {\bf 72}, 043602 (2005).
\bibitem{Frahm}  H. Frahm and G. Palacios, cond-mat/0507368.
\bibitem{Lai}C.K. Lai and C.N.Yang, Phys. Rev A  {\bf 3},
393(1971); C.K.Lai Journ. of Math. Phys., 15, 954(1974).
\bibitem{Batchelor}  M.T. Batchelor, M. Bortz, X.W. Guan, N. Oelkers, cond-mat/0506478.
\bibitem{Yang67} C. N. Yang,  Phys. Rev. Lett. {\bf 19}, 1312(1967).
\bibitem{Gaudin}M. Gaudin, Phys. Lett. A {\bf 24}, 55(1967).
\bibitem{babooks} M. Gaudin, {\it La Fonction d'Onde de Bethe}(Paris, Masson, 1983); F. H. L. Essler {\it et al.}, {\it The One-Dimensional Hubbard Model} (Cambridge University Press, Cambridge, 2005).
\bibitem{takahashi}M. Takahashi, {\it Thermodynamics of one-dimensional solvable models} ( Cambridge University Press, 1999).
\bibitem{Sutherlandbook}B. Sutherland, {\it Beautiful models}(World Scientific Publishing, 2004).
\bibitem{Tsvelick}A.M. Tsvelick and P.B. Wiegmann, Adv. Phys., {\bf 32}, 453(1983).
\bibitem{Andrei}N. Andrei, K. Furuya and J. H. Lowenstein, Rev. Mod. Phys. {\bf 55}, 331(1983).
\bibitem{Sutherland68} B. Sutherland, Phys. Rev. Lett. {\bf 20}, 98 (1968).
\bibitem{LL}E.H. Lieb and W. Liniger, Phys. Rev. {\bf 130}, 1605 (1963);
E.H. Lieb, {\it ibid.}  {\bf 130}, 1616 (1963).
\bibitem{girardeau} M. Girardeau, J. Math. Phys.  {\bf 1}, 516 (1960).
\bibitem{McGuire}J.B.McGuire, J. Math. Phys. {\bf 6}, 432 (1965); {\it ibid}, {\bf 7}, 123 (1966).
\bibitem{ShlyapnikovLDA} K. V. Kheruntsyan {\it et al.},  Phys. Rev. A {\bf 71}, 053615 (2005).
\bibitem{Moritz}H. Moritz {\it et al.}, Phys. Rev. Lett. {\bf 91}, 250402 (2003).
\bibitem{3Dmodes} M. Bartenstein {\it et al.}, Phys. Rev. Lett. {\bf 92}, 203201 (2004)
\bibitem{3dastr}  G. E. Astrakharchik, cond-mat/0507711.
\bibitem{Stringari}S. Stringari,  Phys. Rev. Lett.{\bf 77}, 2360 (1996).
\bibitem{Menotti}S Stringari and  C. Menotti, Phys. Rev. A {\bf 66}, 043610 (2002).
\bibitem{Astrakharchik} G.E.  Astrakharchik {\it et al}, Phys. Rev. Lett. {\bf 93}, 050402
(2004).
\bibitem{bfsumrule}T. Miyakawa, T. Suzuki and H. Yabu, Phys. Rev. A {\bf62}, 063613 (2000).
\bibitem{Lenard}A. Lenard, J. Math. Phys. 5, 930(1964).
\bibitem{Carmelo}J.M.P. Carmelo, cond-mat/0405411.
\bibitem{Woynarovich} F. Woynarovich, J. Phys. C {\bf 15}, 85 (1982).
\bibitem{OgataShiba} M. Ogata and H. Shiba, Phys. Rev. B {\bf 41}, 2326 (1990).
\bibitem{Bragg}S. Richard {\it et al.}, Phys. Rev. Lett. {\bf 91}, 010405 (2003).
\bibitem{Tsvelik} A.M. Tsvelik, {\it Quantum Field Theory in Condensed Matter Physics}(Cambridge University Press, 2003).
\bibitem{Cazalilla}M. A. Cazalilla, Journal of Physics B:  AMOP 37, S1-S47 (2004).
\bibitem{CZ}V.V. Cheianov and M.B. Zvonarev, Phys. Rev. Lett. {\bf 93}, 176401 (2004);
V.V. Cheianov and M.B. Zvonarev, J. Phys. A: Math. Gen. {\bf 37},
2261 (2004);
\bibitem{CZS}V.V. Cheianov, H. Smith and M.B. Zvonarev, Phys.Rev. A {\bf 71}, 033610 (2005).
 \bibitem{BalentsFiete}G.A. Fiete and L. Balents, Phys. Rev. Lett. { \bf 93}, 226401 (2004);
 G.A. Fiete, K.L. Hur and L. Balents, cond-mat/0505186.
 \bibitem{Matveev} K.A. Matveev, Phys. Rev. Lett. {\bf 92}, 106801 (2004);
 K.A. Matveev, Phys. Rev. B {\bf 70}, 245319 (2004).
 \bibitem{Smirnov} V. I. Smirnov, {\it A course of higher mathematics}, Vol IV, p. 24 (Pergamon, Oxford, 1964).
 \bibitem{KBI}V.E. Korepin, N.M. Bogoliubov and A.G. Izergin, {\it Quantum Inverse Scattering Method and Correlation Functions}
 (Cambridge University Press,Cambridge, England,1993).
 \bibitem{IzerginPronko}Izergin A G and Pronko A G  Nucl. Phys. B {\bf 520}, 594 (1998).
 \bibitem{Greiner1d}M. Greiner {\it et al.}, Phys. Rev. Lett. {\bf 87}, 160405 (2001).
 \bibitem{chips}W. Hansel {\it et al.}, Nature {\bf 413}, 498 (2001);
 H. Ott {\it et al.}, Phys. Rev. Lett. {\bf 87} 230401 (2001);
  S. Groth {\it et al.}, Applied Physics Letters  {\bf 85}, 2980 (2004);
 J. Esteve {\it et al.}, Phys. Rev. A {\bf 70}, 043629 (2004);
 S. Aubin {\it et al.}, Journal of Low Temperature Physics {\bf 140}, 377 (2005).
 \bibitem{BECinbox} T.P. Meyrath {\it et al.}, Phys. Rev. A  {\bf 71}, 041604(R) (2005).
 \bibitem{Cote} R. Cote {\it et al.}, Phys. Rev. A {\bf 57}, R4118(1998) .
\bibitem{Yb} K. Honda {\it et al.},  Phys. Rev. A {\bf 66}, 021401(R) (2002);
 Y. Takasu {\it et al.}, Phys. Rev. Lett. {\bf91}, 040404 (2003);
 C. Y. Park and T. H. Yoon , Phys. Rev. A {\bf 68}, 055401 (2003).
\bibitem{Rb} J.P. Burke and J.L. Bohn, Phys. Rev. A {\bf 59}, 1303(1999);
S. G. Crane {\it et al.}, Phys. Rev. A{\bf  62}, 011402(R) (2000).
\bibitem{Olshanii98} M.Olshanii, Phys. Rev. Lett. {\bf 81}, 938(1998).
\bibitem{K41BEC} G. Modugno {\it et al.}, Science 294, 1320 (2001).
\bibitem{Recati}A. Recati {\it et al.}, Phys. Rev. Lett. {\bf 90}, 020401 (2003).
\bibitem{Fuchs}J.N. Fuchs {\it et al.}, cond-mat/0507513.
\bibitem{Gerbier} F. Gerbier {\it et al.}, Phys. Rev. A {\bf 67}, 051602(R) (2003).
\end{thebibliography}
\end{document}